\documentclass{aa}

\usepackage{graphicx}
\usepackage{txfonts}
\usepackage{orcidlink}
\usepackage{hyperref}
\usepackage{booktabs}

\begin{document}

\title{PhoPS: An automated photometric pipeline for survey-era astronomy}

\author{O.~Erece\orcidlink{0000-0002-9723-6823}\inst{1,2}\email{orhane@trgozlemevleri.gov.tr}
\and
Y.~Kilic\orcidlink{0000-0001-8641-0796}\inst{3,4}\corrauth{ykilic@iaa.es}}

\institute{
Türkiye National Observatories, TUG, 07070 Antalya, Türkiye
\and
The Scientific and Technological Research Council of Türkiye (TÜBİTAK), 06680 Ankara, Türkiye
\and
Instituto de Astrofísica de Andalucía (IAA-CSIC), Glorieta de la Astronomía s/n, 18008 Granada, Spain
\and
LIRA, CNRS UMR8254, Observatoire de Paris, Meudon, France
}

\date{Received 23 March, 2026 / Accepted 27 July, 2026}

\abstract
{Modern astronomical surveys generate large volumes of data and provide precise astrometric and photometric reference catalogues. Efficient use of these datasets requires automated, robust, and instrument-independent reduction pipelines.}
{We introduce \texttt{PhoPS} (Photometry and Astrometry of Point Sources), a fully automated Python-based pipeline for photometric reduction with integrated astrometric calibration. It supports homogeneous reduction of stellar and moving Solar System objects, independent of telescope aperture or detector type.}
{\texttt{PhoPS} performs astrometric calibration by dynamically generating local Gaia Data Release 3 (DR3) reference index files propagated to the epoch of observation, avoiding the need for a pre-installed index collection. For photometry, it adopts a field-dependent calibration strategy in which the zero point (ZP) is modelled across the detector plane using Random Sample Consensus (RANSAC)-based linear regression, thereby accounting for spatial systematics such as vignetting and detector non-uniformities.}
{The astrometric contribution was evaluated by comparing solutions obtained with epoch-propagated and non-propagated local Gaia DR3 index files. Using 141\,672 matched measurements from 840 images obtained with the 1-m TUG100 telescope, the propagated solution improves the clipped $N$-weighted total root-mean-square (RMS) residual from 0.284\arcsec\ to 0.241\arcsec, a 15.0\% improvement. The photometric uncertainty model was validated using 203 reference stars and 92\,980 measurements. The normalised residuals are centred close to zero. However, the residual width is magnitude dependent: the limited bright-star bin ($10 \leq G < 13$) shows excess scatter with $\sigma_z=2.069$, the intermediate bin ($13 \leq G < 16$) is closest to unit variance with $\sigma_z=1.129$, and the faintest bin ($16 \leq G < 18$) is conservative with $\sigma_z=0.698$. Thus, the validation reveals a magnitude-dependent uncertainty behaviour rather than a globally consistent error model.}
{\texttt{PhoPS} provides an open-source, lightweight, cross-platform solution for astronomical image reduction. It is suited to asteroid light-curve analysis and stellar variability studies and can diagnose systematics related to telescope tracking, focus stability, and vignetting.}


\maketitle
\nolinenumbers

\section{Introduction} 

The rapid expansion of large-scale astronomical surveys over the past decade has significantly transformed observational astronomy. Ground-based programmes such as Two Micron All Sky Survey (2MASS; \citealt{2mass}), Sloan Digital Sky Survey (SDSS; \citealt{sdss}), and the Vera C. Rubin Observatory Legacy Survey of Space and Time (LSST; \citealt{lsst}) have significantly enriched our view of the sky, providing catalogues containing billions of sources. Together with ground-based observatories, space-based missions have further increased the level of accuracy. While Hipparcos \citep{hipparcos} established the first high-quality, all-sky astrometric framework, its successor, Gaia \citep{Gaia2016b, Gaia2016a, Gaia2018, Gaia2023}, has had a significant impact on the field. By providing high-precision astrometric and multi-band photometric measurements for nearly two billion sources, Gaia now delivers an extremely precise astrometric reference frame that serves as the standard for celestial calibration.

The large number of Gaia DR3 sources ensures that most astronomical fields contain a sufficient number of well-characterised reference stars. This abundance allows for a direct astrometric plate solving between raw pixel coordinates and the International Celestial Reference System (ICRS), as well as magnitude transformation to standard photometric systems. However, effectively utilising the potential of such a large amount of data requires algorithms capable of efficiently querying, handling, and exploiting source information for tasks such as plate solving, source matching, magnitude transformation, and the characterisation of instrumental effects (e.g., field-dependent sensitivity and optical distortions).

While professional observatories develop dedicated and highly optimised reduction pipelines (e.g. \citealt{esoreflex, edps, dragons, lsst, panstarrs}), these systems are typically developed for specific instruments or surveys and are therefore not always straightforward to apply to heterogeneous datasets. Examples of more portable alternatives include open-source Astropy-based workflows \citep{astropy:2013, astropy:2018, astropy:2022}, \texttt{STDPipe} \citep{karpov2021}, \texttt{PHOTOMETRYPIPELINE} \citep{mommert2017}, and \texttt{PRAIA} \citep{assafin2023}. These tools differ in design philosophy, particularly in whether they are tied to dedicated infrastructures or intended to be portable, whether they are instrument-specific or more broadly applicable, and in their astrometric indexing and photometric calibration strategies. \texttt{STDPipe} is a modular set of Python routines for astrometry, photometry, and transient-detection-related tasks, intended for the implementation of custom pipelines as well as interactive data analysis. \texttt{PHOTOMETRYPIPELINE}, by contrast, is a more explicitly end-to-end automated pipeline aimed at calibrated point-source photometry, particularly for small- to medium-sized observatories and moving target observations. \texttt{PRAIA} is a mature package for the automatic reduction of heterogeneous astronomical images, with a strong emphasis on high-precision astrometry for Solar System applications, including natural satellites, near-Earth asteroids, trans-Neptunian objects, and stellar occultation predictions.

In this context, the emphasis of \texttt{PhoPS} is complementary and is not intended to replace these existing tools: it combines portable Gaia-based astrometric index generation with field-dependent photometric calibration in a single automated workflow. In particular, its use of dynamically generated local Gaia DR3-based astrometric indexes and a spatially dependent zero point (ZP) model distinguishes it from workflows based primarily on static astrometric references or a single global photometric zero point.

On the other hand, the development and maintenance of IRAF\footnote{\url{https://iraf-community.github.io/}} \citep{iraf:1986, iraf:1993}, once one of the most widely used tools for astronomical image reduction, was discontinued by the National Optical Astronomy Observatories (NOAO) in 2013. Following this, NOAO recommended that new projects adopt alternative solutions such as Astropy. Although IRAF was highly capable for data reduction tasks, it was not well suited to large-scale automated pipelines \citep{panstarrs}. Many of the algorithms originally implemented in IRAF are still referenced by newly developed software, including the present work. Together with the widespread adoption of the Python programming language and significant recent developments in Astropy, there are many ongoing efforts to develop both global and instrument-specific data processing pipelines.

As astronomical imaging is increasingly performed across a wide range of applications, not only in astrophysics but also in fields such as space surveillance, there is an increasing demand for flexible and automated tools for astrometry and photometry. In this context, integrated and portable workflows that combine astrometric calibration with field-dependent photometric modelling remain comparatively uncommon, especially for heterogeneous datasets obtained with different optical systems. Such tools must be automated, scalable, and robust enough to support data from small-aperture telescopes and robotic systems to large-scale professional observatories.

In this paper, we introduce \texttt{PhoPS} (Photometry and Astrometry of Point Sources), an open-source and cross-platform software package designed to provide consistent and precise astrometric and photometric reductions. \texttt{PhoPS} is suitable for both stellar and Solar System objects. It supports astrometric and photometric measurements of any detectable point source within the image, such as stellar sources or asteroids. Although different databases can be utilised, \texttt{PhoPS} uses, by default, the Gaia Data Release 3 (DR3) catalogue, not only as a static reference catalogue but also as a dynamic source for local astrometric index generation and field-dependent photometric zero-point modelling. Gaia DR3 provides a comprehensive set of stellar parameters, including coordinates, proper motions, parallaxes, and multi-band photometry, making it well suited for this purpose.

In the following sections, we describe the overall architecture and workflow of \texttt{PhoPS} (Sect.~\ref{sec:software}), present validation results (Sect.~\ref{sec:validation}), discuss the implications of the algorithm's field-dependent calibration (Sect.~\ref{sec:discussion}), and summarise our conclusions (Sect.~\ref{sec:conclusion}).

\section{Software description and algorithm}\label{sec:software}

\subsection{Overview of \texttt{PhoPS}} \label{sec:software_overview}

\texttt{PhoPS} is an automated data reduction pipeline developed for astrometric and photometric analysis of astronomical images in Flexible Image Transport System (FITS) format \citep{fits} and is publicly available online\footnote{\url{https://github.com/orhanerece/PhoPS}}, where the source code, documentation, and usage examples are provided. The pipeline is written in Python, making it cross-platform, and relies on several open-source libraries, including Astropy and its coordinated packages Photutils \citep{photutils} and Astroquery \citep{astroquery}. Source detection is performed using the \texttt{DAOStarFinder} algorithm \citep{Daofind}, while astrometric solutions are obtained through Astrometry.net \citep{astrometry}. By combining these widely used tools within a unified framework, \texttt{PhoPS} performs both astrometric and photometric reductions in an automated workflow.

One of the main objectives of \texttt{PhoPS} is to derive both astrometric and photometric solutions using Gaia DR3 sources as the reference frame. Using a single reference catalogue for both calibrations simplifies the workflow and keeps the pipeline lightweight. \texttt{PhoPS} is organised in a modular way, where key processing steps such as astrometry, photometry, and target identification are implemented as independent modules and controlled through a configuration file.

The pipeline operates in two different modes, depending on the type of target:

\begin{itemize}

\item In \texttt{star} mode, the pipeline performs standard reductions for stellar sources. After the World Coordinate System (WCS) solution is obtained, targets are identified using their right ascension (RA) and declination (Dec) rather than fixed pixel positions. This approach allows the target to be located even if there are offsets between images.

\item In \texttt{asteroid} mode, the pipeline queries the JPL Horizons\footnote{\url{https://ssd.jpl.nasa.gov/horizons/app.html}} service to obtain ephemerides corresponding to the mid-exposure Julian Date (JD) and the observer’s location for each image. These ephemerides are then used to locate moving Solar System objects in the images. The subsequent reduction steps are identical to those used in the star mode.

\end{itemize}

\begin{figure*} \centering \includegraphics[width=\linewidth]{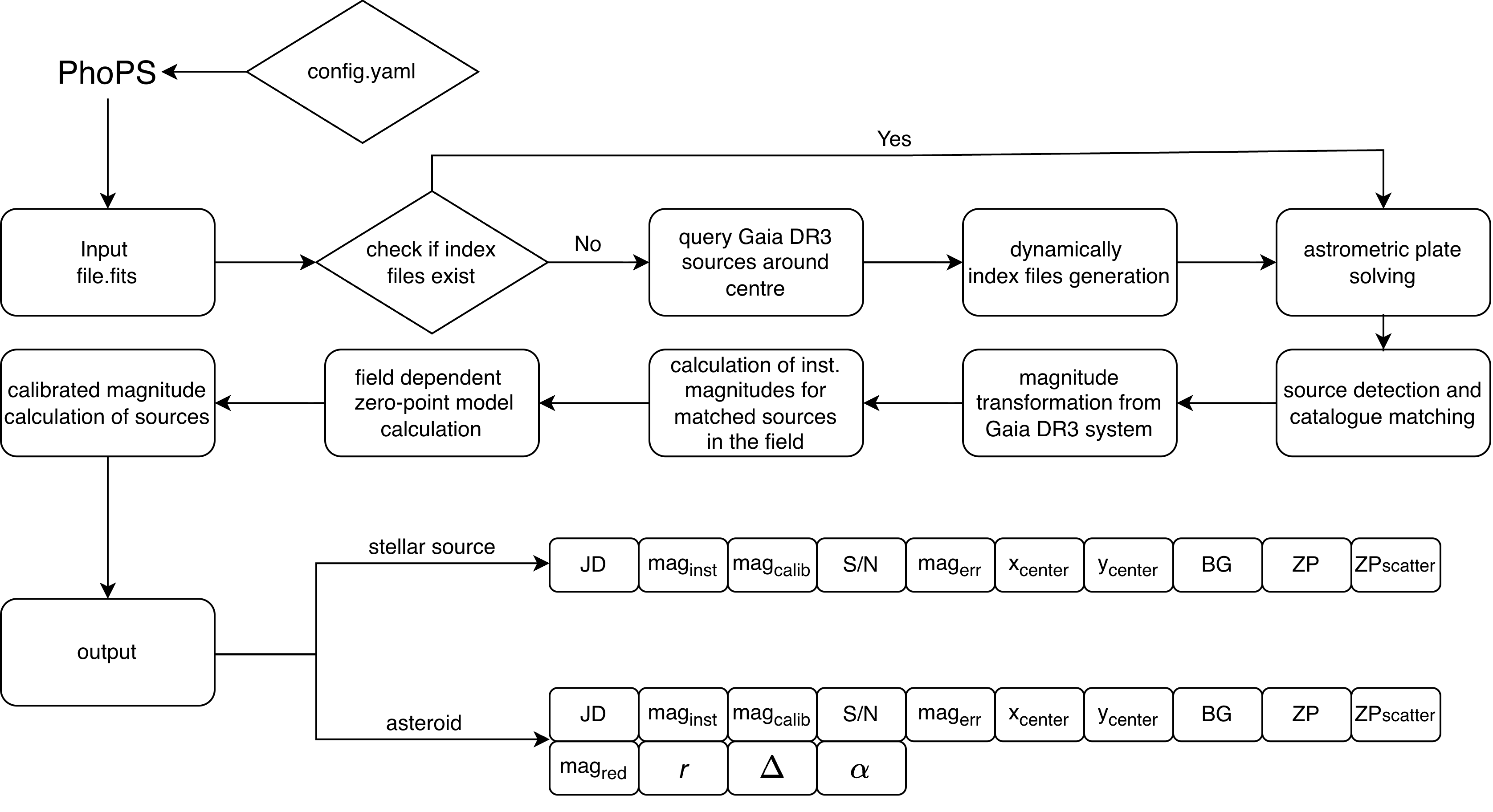} \caption{Workflow of \texttt{PhoPS}. JD denotes the Julian date;
$\mathrm{mag}_{\rm inst}$, instrumental magnitude;
$\mathrm{mag}_{\rm calib}$, calibrated magnitude;
S/N, signal-to-noise ratio;
$\mathrm{mag}_{\rm err}$, total photometric uncertainty;
$x_{\rm center}$ and $y_{\rm center}$, physical coordinates of the target;
BG, background value;
ZP, zero-point magnitude;
$\mathrm{mag}_{\rm red}$, reduced magnitude of the asteroid;
$r$, heliocentric distance of the asteroid;
$\Delta$, geocentric distance of the asteroid;
and $\alpha$, phase angle of the asteroid.}
\label{fig:phops_diagram} \end{figure*}

The pipeline is designed to operate across a wide range of instruments and observing configurations and can be applied to any FITS-compliant dataset. The overall workflow of the algorithm is illustrated in Fig.~\ref{fig:phops_diagram}.

\subsection{Dynamic astrometric index generation} \label{sec:dynamicindex} 

\texttt{PhoPS} uses the Astrometry.net \citep{astrometry} software suite for astrometric calibration. Traditional implementations of Astrometry.net rely on locally stored, static index files containing positional patterns derived from reference catalogues. These files often occupy significant storage and must be pre-matched to the specific field of view (FoV) of the instrument, limiting the portability of the software. To overcome these limitations, \texttt{PhoPS} introduces a dynamic workflow that generates localised reference indexes on-the-fly. In this workflow, \texttt{PhoPS} prepares a local Gaia DR3-based reference dataset propagated to the epoch of observation, and then orchestrates the subsequent index-generation and plate-solving steps using the \texttt{hpsplit}, \texttt{build-astrometry-index}, and \texttt{solve-field} commands of Astrometry.net. Thus, \texttt{PhoPS} extends the standard Astrometry.net workflow by dynamically constructing field-specific reference inputs from Gaia DR3, while the index construction, pattern matching, and WCS derivation are performed by Astrometry.net.

The process begins by extracting the central coordinates and FoV geometry from the FITS header. Using these parameters, the pipeline issues an asynchronous Astronomical Data Query Language (ADQL) query to the Gaia DR3 archive. This query retrieves essential parameters, including magnitudes, $G_{\mathrm{BP}}-G_{\mathrm{RP}}$ colours, coordinates, proper motions, and parallaxes, ordered by the brightness of sources. While Gaia DR3 datasets are at the epoch J2016.0, \texttt{PhoPS} propagates the sources using proper motions and parallaxes (neglecting radial velocity when unavailable) to the desired epoch to ensure high astrometric accuracy. The retrieved data are then stored locally as a FITS file, hereafter referred to as the Gaia reference file (\texttt{gaia\_reference\_\{unique\_id\}.fits}). This file is used for multiple purposes: for temporary astrometric index generation, for source crossmatching (see Sect. \ref{sec:crossmatch}), and for magnitude transformation (see Sect. \ref{sec:photocal}).

Generating index files starts by selecting a spatially uniform and relatively bright subset of stars from the Gaia reference file. The data are then divided into equal-area HEALPix tiles \citep{healpix}, with a resolution defined by the \texttt{NSIDE} parameter, using the \texttt{hpsplit} utility. This ensures a uniform spatial sampling of reference stars across the field. The tiles are constructed with overlapping boundaries to allow reliable pattern matching for sources located near tile edges. Subsequently, the \texttt{build-astrometry-index} command is run for each tile across a range of quad scales defined in the configuration file. To allow for the simultaneous storage of multiple observation fields without collision, each generated index is assigned a unique identifier and filename suffix based on its celestial coordinates. Thus, index files can be reused for subsequent images covering the same field and requiring the same epoch-propagated reference solution, reducing repeated archive queries and avoiding unnecessary regeneration of local reference inputs within a given observing sequence. Finally, the \texttt{solve-field} utility of Astrometry.net uses these index files to perform the blind plate-solving step and derive the WCS solution. Once a solution is successful, the reference pixels and related values are written directly into the FITS header. 

In the present workflow, the primary motivation for constructing local astrometric indexes dynamically is to build them from Gaia DR3 source positions propagated to the epoch of each observation. Since Gaia DR3 astrometry is referenced to epoch J2016.0, neglecting proper motions can introduce non-negligible positional offsets in datasets acquired several years away from the catalogue epoch, especially for narrow-field instruments and multi-epoch observations. By querying the relevant field directly and propagating the source positions before index construction, \texttt{PhoPS} provides an epoch-consistent local astrometric reference for the subsequent Astrometry.net workflow.

A practical advantage of this approach is that the pipeline remains portable and self-contained, requiring no permanent pre-installed astrometric index collection beyond the images being processed. Once the approximate pointing and field of view are obtained from the FITS header, or supplied by the user when necessary, the required local Gaia DR3-based reference inputs can be generated on demand on any machine. Thus, this strategy is not presented as a speed or storage advantage over a complete precomputed Astrometry.net installation, but as a way to provide an epoch-consistent and portable astrometric reference construction based on Gaia DR3 positions propagated to the observation epoch. The locally generated indexes are stored and can be reused for subsequent observations of the same field, which provides an additional practical benefit for repeated time-series reductions.

\subsection{Source extraction and Gaia-based crossmatching} \label{sec:crossmatch}
Following a successful astrometric solution, \texttt{PhoPS} performs source extraction on the WCS-calibrated image. Source detection is carried out using the \texttt{DAOStarFinder} implementation of the \texttt{DAOFind} algorithm \citep{Daofind} available in the Photutils package, which is widely used for point-source detection in stellar fields. The detection parameters are defined in the configuration file, most importantly the detection threshold and the expected full width at half maximum (FWHM) of stellar profiles. Although a relatively low threshold increases the number of initial source detections, subsequent crossmatching and quality filtering remove spurious sources before photometric calibration. Once the sources are detected, their centroid positions are refined using a two-dimensional quadratic centroiding method described by \cite{centroid_quadratic}. The refined source positions are then crossmatched with sources in the Gaia reference file.

To obtain a clean set of reference sources within the image, the raw detections are filtered in several stages:

\begin{enumerate}

\item Saturation and edge filtering:
Detections exceeding the detector linearity limit are removed to avoid unreliable centroid and photometric measurements. Sources located close to the physical edges of the charge-coupled device (CCD) are also discarded, as aperture measurements may be incomplete and the photometric response can be less reliable near the detector boundaries.

\item Positional crossmatching:
The remaining detections are crossmatched with the Gaia reference file generated locally during the astrometric stage (Sect.~\ref{sec:dynamicindex}). The matching is performed using the \texttt{SkyCoord} functionality of Astropy, which computes the two-dimensional angular separation between sources. A default matching radius of 2 arcseconds is adopted, which is appropriate given the typical astrometric precision and pixel scales of the datasets considered. Because the Gaia catalogue subset is stored locally, the crossmatching is performed without additional archive queries, improving computational efficiency.

\item Isolation and quality control:
An isolation criterion is applied to reduce flux contamination in crowded regions. Any detected source with a neighbouring object within a user-defined radius (typically 10 pixels) is excluded from the reference sample. In addition, only Gaia sources with measured $G$-band magnitudes and available $G_{\mathrm{BP}}-G_{\mathrm{RP}}$ colours are used since both quantities are required for the photometric calibration described in Sect.~\ref{sec:photocal}.

\end{enumerate}

\begin{figure}
\centering \includegraphics[width=\linewidth]{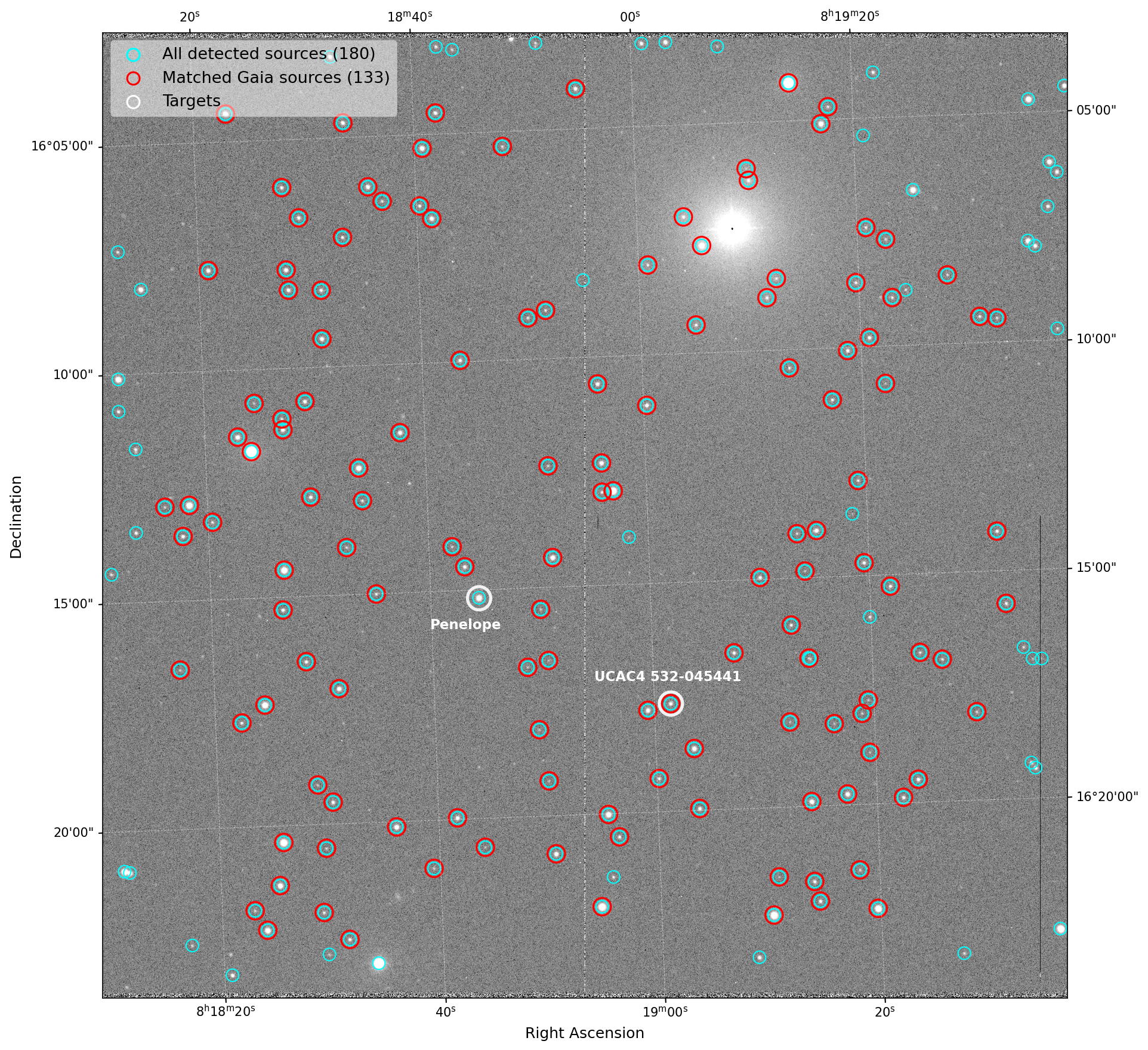} \caption{An example of a frame demonstrating the automated source detection and Gaia DR3 crossmatching performance of \texttt{PhoPS}. Blue circles mark the 180 sources detected by \texttt{DAOStarFinder} above the adopted detection threshold, while red circles mark the 133 sources successfully matched to the local Gaia DR3 reference catalogue. The target objects, asteroid (201) Penelope and UCAC4 532-045441, are indicated by white circles.}
\label{fig:sourcematch}
\end{figure}

After these filtering steps, \texttt{PhoPS} produces a list of unsaturated and well-isolated crossmatched reference stars suitable for photometric calibration (see Fig.~\ref{fig:sourcematch}). The calculation of the photometric zero-point model is described in the next section.

\subsection{Photometric calibration}\label{sec:photocal}

The photometric calibration begins by converting the Gaia $G$ magnitudes of the crossmatched reference stars to the filter of the image. \texttt{PhoPS} performs this transformation using the equation:

\begin{equation}
m_{\mathrm{ref}} = G - \sum_{i=0}^{p} a_{i}\,(G_{\mathrm{BP}} - G_{\mathrm{RP}})^{i},
\label{eq:magtransform}
\end{equation}

where $m_{\mathrm{ref}}$ is the transformed magnitude of a reference star in the observed passband, $G$ is the Gaia DR3 magnitude, $G_{\mathrm{BP}} - G_{\mathrm{RP}}$ is the colour, and the coefficients $a_i$ are taken from \cite{gaia_mag_transform2022} for the specific filter corresponding to the dataset.

After transforming the reference stars to the appropriate photometric system, we need to measure their instrumental magnitudes directly from the image. \texttt{PhoPS} performs circular aperture photometry on all sources using the Photutils package. The user specifies the aperture size in the configuration file, choosing between a fixed pixel radius, a fixed angular radius in arcseconds, or a scaling factor applied to the stellar FWHM. When the FWHM-factor method is selected, \texttt{PhoPS} calculates a representative FWHM for the image and multiplies it by the given factor to determine the aperture radius. Local background is estimated via a sigma-clipped median within an annulus around each source, and the background-subtracted flux $F$ yields the instrumental magnitude as $m_{\mathrm{inst}} = -2.5 \log_{10}(F)$. The resulting instrumental magnitudes and propagated formal photometric uncertainties, $\sigma_{\rm formal}$, are retained for both the science target and the reference stars, allowing the reference-star population to be used later for time-series validation of the photometric uncertainty model. Atmospheric extinction is not explicitly modelled and is assumed to be approximately constant across the field for a given exposure, such that its effect is absorbed into the zero-point term. This assumption is generally valid for small fields of view and short exposure times.

For each reference star, we define the individual zero point as the difference $\Delta m = m_{\mathrm{ref}} - m_{\mathrm{inst}}$. If the detector were perfectly flat and all reference stars standard, these $\Delta m$ values would be constant across the image. In practice, they show spatial structure due to vignetting, detector non-uniformities, and intrinsic variability in some sources. To see how $\Delta m$ behaves across the field, \texttt{PhoPS} models it as a function of the radial distance $r$ from the detector centre $(x_0, y_0)$:
\begin{equation}
r = \sqrt{(x - x_0)^2 + (y - y_0)^2}.
\end{equation}
We adopt a linear model for this radial dependence:
\begin{equation}
f(r \mid \theta) = \alpha + \beta r,
\label{eq:linearmodel}
\end{equation}
where $\theta = (\alpha, \beta)$ is the vector of fitted parameters. However, the sample may contain outliers due to astrophysical and instrumental effects, such as variable stars, bad pixels, and crossmatching errors, which would bias a standard least-squares fit.

\begin{figure}
\centering \includegraphics[width=\linewidth]{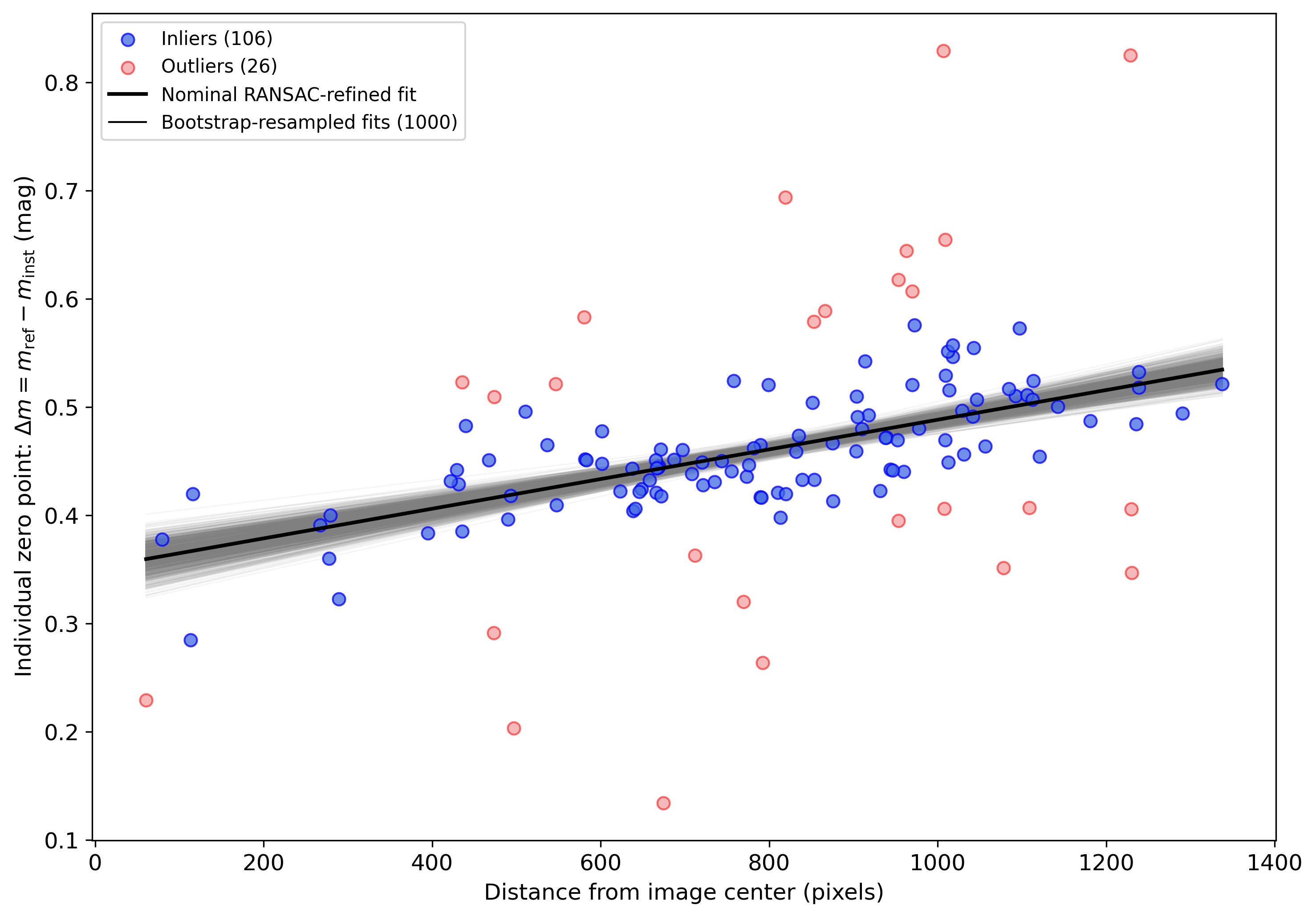} 
\caption{Photometric zero-point modelling for a representative frame. Blue points show the final RANSAC inlier set, while red points denote rejected outliers. The solid black line represents the nominal RANSAC-refined radial zero-point model. Thin grey lines show linear zero-point models obtained from bootstrap resampling of the final inlier set, illustrating the uncertainty of the field-dependent zero-point fit.}
\label{fig:ransac}
\end{figure}

To mitigate this, we employ the Random Sample Consensus (RANSAC) algorithm \citep{ransac}. RANSAC iteratively identifies a robust inlier set, $\hat{\mathcal{I}}$, by selecting the largest subset of data points consistent with the assumed model within a defined residual threshold and over a fixed number of iterations. In the general pipeline configuration, the RANSAC residual threshold can either be specified by the user or calculated automatically from the data. In the automatic mode, \texttt{PhoPS} tests a set of candidate thresholds and records the number of inliers obtained for each value. The final threshold is chosen using a knee criterion applied to the threshold--inlier-count relation. This provides a practical balance between rejecting outliers and retaining a sufficient number of useful reference stars for the radial zero-point fit. For time-series data, the threshold is determined from the first valid frame and then kept fixed for the remaining frames, ensuring a homogeneous rejection criterion throughout the sequence.

\begin{equation}
\tilde{\theta} = \arg\min_{\theta} \sum_{i \in \hat{\mathcal{I}}} \bigl[ \Delta m_{i} - f(r_{i} \mid \theta) \bigr]^{2}.
\end{equation}
The radial distance-dependent zero-point magnitude is then defined as $ZP(r) = f(r \mid \tilde{\theta})$, which characterises the spatial variation of the zero point across the detector plane. This RANSAC-based radial modelling effectively accounts for large-scale systematic variations across the field, such as field-dependent sensitivity changes due to optical vignetting (see Fig.~\ref{fig:ransac}). For a target object located at a radial distance $r_{\mathrm{target}}$ from the detector centre, the calibrated magnitude in the adopted photometric system is finally computed as:

\begin{equation}
m_{\mathrm{standard}} = m_{\mathrm{inst, target}} + ZP(r_{\mathrm{target}}).
\label{eq:calibmag}
\end{equation}
This field-dependent calculation of zero-point magnitude depends on where the target is located on the detector, ensuring accurate results instead of using a single zero-point magnitude for the entire FoV. The uncertainty associated with the field-dependent zero point is estimated by bootstrap resampling of the final RANSAC inlier set, $\hat{\mathcal{I}}$. 
In each bootstrap realisation, the inlier pairs $(r_i,\Delta m_i)$ are randomly resampled with replacement, and the linear radial model is fitted again. This produces a set of bootstrap zero-point models:
\begin{equation}
ZP^{(k)}(r) = \alpha^{(k)} + \beta^{(k)} r,
\label{eq:zpbootstrap}
\end{equation}
where $k$ denotes the bootstrap realisation. The adopted zero-point correction used in Eq.~\ref{eq:calibmag} is obtained from the nominal RANSAC-refined fit to the full final inlier set. The bootstrap realisations provide a distribution of zero-point estimates at each source position, which is used to quantify the uncertainty of the adopted correction. For a source located at radial distance $r_j$, this uncertainty is defined as

\begin{equation}
\sigma_{ZP,{\rm boot}}(r_j) =
{\rm std}_{k}\left[ZP^{(k)}(r_j)\right].
\label{eq:zpbooterr}
\end{equation}
The total photometric uncertainty reported by \texttt{PhoPS} is then computed by adding the formal aperture-photometry uncertainty and the bootstrap zero-point uncertainty in quadrature:
\begin{equation}
\sigma_{{\rm total},j}
=
\left[
\sigma_{{\rm formal},j}^{2}
+
\sigma_{ZP,{\rm boot}}^{2}(r_j)
\right]^{1/2}.
\label{eq:sigmatotal}
\end{equation}
The resulting total uncertainty, $\sigma_{\rm total}$, is reported as \texttt{mag\_err} for both the science target and the reference-star measurements, which are used for the population-level validation presented in Sect.~\ref{sec:validation}.

\section{Validation and results}\label{sec:validation}
We performed a series of validation tests to assess the astrometric and photometric performance of \texttt{PhoPS}. The analysis focuses on (i) astrometric residuals relative to the Gaia DR3 reference frame, (ii) the stability of the spatially dependent photometric zero-point model, and (iii) the scientific quality of the resulting light curves for both stellar and moving targets.

\subsection{Observations and dataset}\label{sec:dataset}

To evaluate the performance of \texttt{PhoPS}, we used datasets obtained with two different optical systems operated at the Türkiye National Observatories, namely the 1-m TUG100 telescope and the 0.6-m TUG060 robotic telescope. In addition, the pipeline has already been applied to data from other facilities in previous studies. The main characteristics of the observing systems on which \texttt{PhoPS} has been used are summarised in Table~\ref{tab:systems}.

\begin{table*}
\caption{Observing systems and datasets used with \texttt{PhoPS}.}
\label{tab:systems}
\centering
\begin{tabular}{lcccccc}
\hline\hline
System & Diameter (m) & Instrument & Format (pixels) &
Pixel scale & Field of view & Studies \\
\hline
TUG100 & 1.0 & SI 1100 & $2048 \times 2048$ &
0.62\arcsec & 21.5\arcmin $\times$ 21.5\arcmin &
This work; previous study\tablefootmark{a} \\

TUG060 & 0.6 & Andor DZ936 & $2048 \times 2048$ &
0.456\arcsec & 15.6\arcmin $\times$ 15.6\arcmin &
This work \\

RTT150 & 1.5 & TFOSC & $2048 \times 2048$ &
0.39\arcsec & 13\arcmin $\times$ 13\arcmin &
Previous study\tablefootmark{a} \\

NTT & 3.58 & EMMI & $1038 \times 2055$ &
0.27\arcsec & 9.1\arcmin $\times$ 8.6\arcmin &
Previous study\tablefootmark{b} \\

VLT UT2 & 8.2 & FORS1 & $2048 \times 1034$ &
0.126\arcsec & 4.30\arcmin $\times$ 2.17\arcmin &
Previous study\tablefootmark{b} \\
\hline
\end{tabular}

\tablefoot{TUG100 values are given for
$2\times2$ binning as used in the referenced studies. The NTT/EMMI and VLT/FORS1 entries correspond
to the image configurations used in the referenced study.
\tablefoottext{a}{\cite{erece2023}.}
\tablefoottext{b}{\cite{kilic2026}.}}
\end{table*}

The dataset used in this study includes time-series imaging obtained with the standard Johnson-Cousins $V$ filter for TUG100 and with the $B$, $V$, $R$, and $I$ filters for TUG060. Before being processed with \texttt{PhoPS}, all raw frames were bias- and dark-subtracted and corrected using flat-field images. The use of data from different instruments allows us to assess the performance of the pipeline under different pixel scales, fields of view, and detector characteristics.

The main quantitative validation presented in this section is based on a TUG100 time series of (201)~Penelope consisting of 840 CCD frames obtained in the Johnson-Cousins $V$ filter. In addition to the moving-target photometry, \texttt{PhoPS} records calibrated measurements for the Gaia-matched reference stars in each frame. This makes it possible to use the same dataset both for representative target light curves and for a population-level validation of the reported photometric uncertainties. The same aperture-photometry configuration was applied consistently to the moving target and to the reference-star measurements throughout the final validation.

The reference-star uncertainty validation was restricted to calibration-inlier measurements in the range $10 \leq G < 18$. Reference stars with fewer than ten valid measurements were excluded. To remove candidate variable or unstable sources, we applied a stability filter based on the ratio between each star's robust calibrated magnitude scatter and its mean reported total uncertainty, evaluated separately within each magnitude bin. Stars with outlying ratios were rejected from the validation sample. The final stability-filtered calibration-inlier sample contains 203 reference stars and 92\,980 individual measurements from 839 frames.

In order to evaluate both the stellar and moving-object reduction capabilities of \texttt{PhoPS}, we considered several representative targets observed with these systems. The main example presented in this section is the asteroid (201)~Penelope, which has a published rotation period of $3.7474 \pm 0.0001$ h (e.g. \citealt{penelope}). In addition, to assess the photometric stability of the reduction within the same (201)~Penelope dataset, we also analyse the field star UCAC4 532-045441 as an internal comparison source. Its catalogued $V$-band magnitude is 15.421 mag\footnote{\url{https://simbad.cds.unistra.fr/simbad/sim-id?Ident=\%4027049349&Name=UCAC4\%20532-045441}} \citep{ucac4star}, and its calibrated light curve is examined together with that of (201)~Penelope in order to test whether the reduction preserves the intrinsic variability of the moving target without introducing correlated trends in a field source. Additional examples based on TUG060 data of the short-period eclipsing binary V523 Cas (e.g. \citealt{v523cas1, v523cas3, v523cas2}) and TUG100 data of asteroid 19184 (1991 TB6) \citep{maria} are provided in Appendix~\ref{sec:appendix}. These supplementary cases are included to illustrate the applicability of the pipeline to both stellar variability and moving-object studies, as well as the main validation example discussed here. An additional example of a frame affected by a transient tracking error case is also presented in Appendix~\ref{sec:appendix}.

\subsection{Astrometric accuracy}\label{subsec:astrometricaccuracy}
The Gaia DR3 catalogue provides comprehensive astrometric solutions for approximately 1.8 billion sources, with uncertainties typically ranging from 0.01 to 1 mas depending on source magnitude \citep{Gaia2021-astrometry1, Gaia-astrometry2}. \texttt{PhoPS} queries the \texttt{gaiadr3.gaia\_source} table via ADQL to obtain Gaia DR3 references, as described in Sect.~\ref{sec:crossmatch}. The astrometric contribution of \texttt{PhoPS} lies in the dynamic generation of Gaia DR3 index files tailored to each observed field, while the actual plate-solving is performed by the \texttt{solve-field} utility of Astrometry.net using these index files. In the present workflow, Simple Imaging Polynomial (SIP) distortion terms were included through the built-in Astrometry.net \texttt{tweak\_order} parameter. Polynomial orders from 2 to 6 were tested, and order 5 was adopted because it produced the lowest astrometric scatter across the dataset used in this analysis. No additional external high-order astrometric refinement step was applied beyond that stage.

To evaluate the contribution of \texttt{PhoPS}'s epoch propagation, we calculated the positional residuals ($\Delta \alpha \cos \delta$ and $\Delta \delta$) between sources detected in the observed frames and the corresponding Gaia DR3 positions under two conditions: (i) using dynamically generated index files with positions propagated to the observation epoch via proper motions and parallax (hereafter ``propagated''), and (ii) using the same dynamically generated index files without epoch propagation, retaining the native J2016.0 Gaia DR3 positions (hereafter ``non-propagated''). This comparison isolates the effect of epoch propagation within \texttt{PhoPS} while keeping the plate-solving engine and index-generation procedure identical in both cases.
To ensure a strictly controlled comparison, only sources present in both solutions were retained. In total, we analysed 141\,672 matched measurements from 840 images obtained with the TUG100 telescope, restricted to the magnitude range $10 \leq G < 18$. The images were taken at different sky positions and covered a range of observing conditions; no images were rejected based on poor tracking or focus errors, providing a realistic assessment of the expected performance under typical observing conditions. Sources were grouped into three magnitude bins --- $10 \leq G < 13$, $13 \leq G < 16$, and $16 \leq G < 18$ --- and root-mean-square (RMS) values were computed after applying a 3$\sigma$ clipping algorithm within each bin to reduce the impact of outliers arising from source mismatches and transient observing artefacts.
The propagated solution yields lower residuals in all magnitude regimes. For bright sources ($10 \leq G < 13$), the RMS values are $0.173\arcsec$ in RA and $0.059\arcsec$ in Dec, corresponding to a total RMS of $0.183\arcsec$, compared with $0.193\arcsec$ in RA, $0.096\arcsec$ in Dec, and $0.215\arcsec$ total for the non-propagated case, representing an improvement of 15.1\%. For intermediate magnitudes ($13 \leq G < 16$), the propagated solution gives $0.161\arcsec$ in RA and $0.068\arcsec$ in Dec, corresponding to a total RMS of $0.175\arcsec$, whereas the non-propagated solution yields $0.186\arcsec$ in RA, $0.126\arcsec$ in Dec, and $0.225\arcsec$ total, corresponding to an improvement of 22.2\%. For fainter sources ($16 \leq G < 18$), the propagated solution gives $0.269\arcsec$ in RA and $0.142\arcsec$ in Dec, corresponding to a total RMS of $0.304\arcsec$, compared with $0.297\arcsec$ in RA, $0.176\arcsec$ in Dec, and $0.345\arcsec$ total in the non-propagated case, corresponding to an improvement of 11.9\%. The clipped N-weighted total RMS over all bins decreases from $0.284\arcsec$ in the non-propagated case to $0.241\arcsec$ in the propagated case, corresponding to an overall improvement of 15.0\%. The magnitude-binned and overall astrometric results are summarised in Table~\ref{tab:astrometric_comparison}.

\begin{table*}
\centering
\small
\caption{Comparison of astrometric residuals obtained with propagated
and non-propagated Gaia DR3 index files.}
\label{tab:astrometric_comparison}
\begin{tabular}{lcccccccc}
\toprule
& & \multicolumn{3}{c}{Propagated} & \multicolumn{3}{c}{Non-propagated} & \\
\cmidrule(lr){3-5} \cmidrule(lr){6-8}
$G$ bin & $N_{\rm clip}$ & RMS$_{\rm RA}$ & RMS$_{\rm Dec}$ & RMS$_{\rm tot}$ & RMS$_{\rm RA}$ & RMS$_{\rm Dec}$ & RMS$_{\rm tot}$ & Improvement in RMS$_{\rm tot}$ \\
\midrule
$10 \leq G < 13$ & 8\,415  & 0.173 & 0.059 & 0.183 & 0.193 & 0.096 & 0.215 & 15.1\% \\
$13 \leq G < 16$ & 68\,182 & 0.161 & 0.068 & 0.175 & 0.186 & 0.126 & 0.225 & 22.2\% \\
$16 \leq G < 18$ & 59\,970 & 0.269 & 0.142 & 0.304 & 0.297 & 0.176 & 0.345 & 11.9\% \\
\midrule
Overall & 136\,567 & 0.216 & 0.107 & 0.241 & 0.241 & 0.149 & 0.284 & 15.0\% \\
\bottomrule
\end{tabular}
\tablefoot{Only measurements of sources common to both
solutions are included. RMS values are given in arcseconds after
$3\sigma$ clipping within each magnitude bin. Percentage improvements
were calculated from the unrounded RMS values.}
\end{table*}

\begin{figure}
\centering
\includegraphics[width=\linewidth]{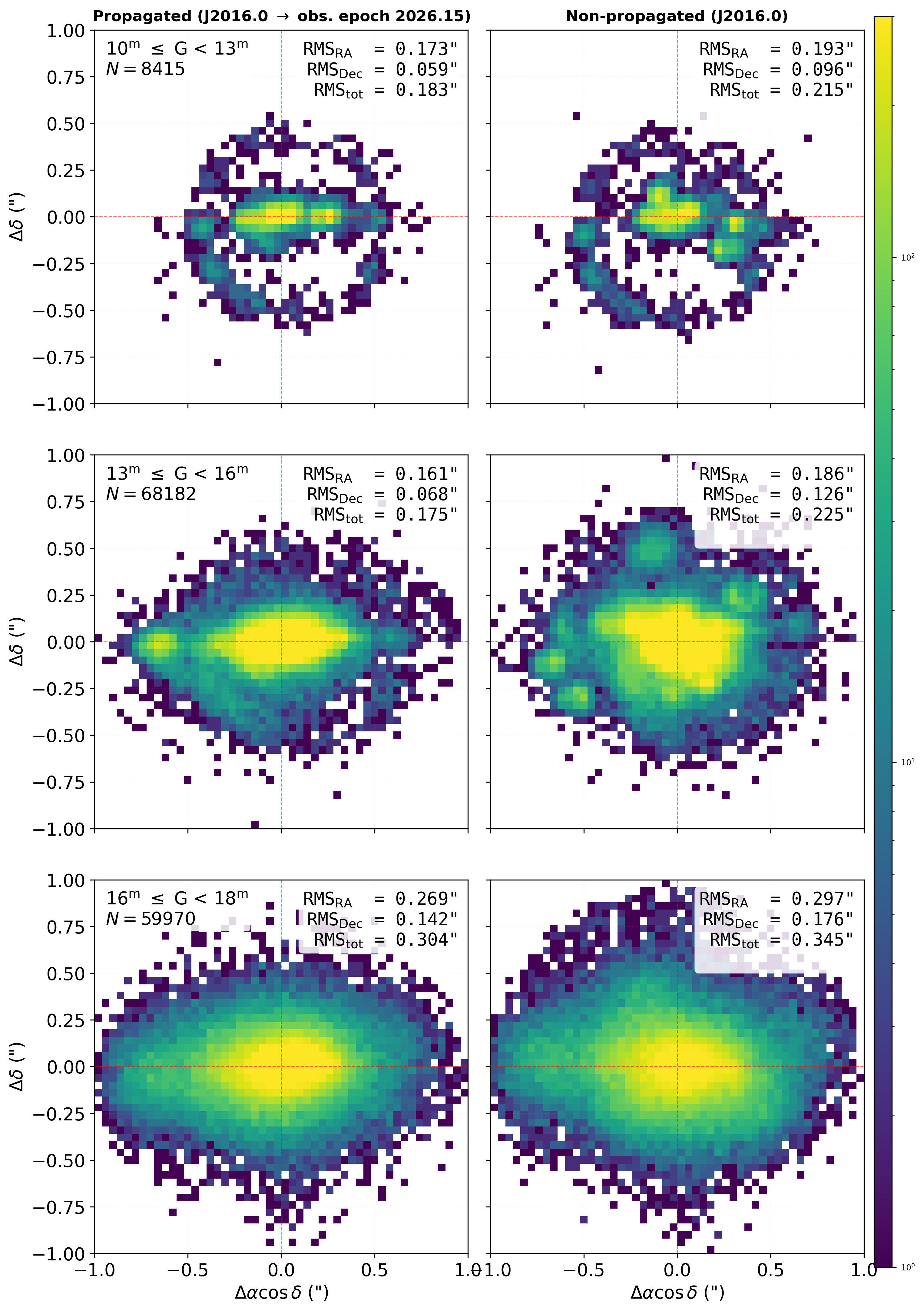}
\caption{Astrometric residual distributions for 141\,672 matched measurements of sources common to both the propagated and non-propagated solutions. Each panel corresponds to a Gaia $G$-magnitude bin, and RMS values are derived after 3$\sigma$ clipping. The colour scale is identical across all panels.}
\label{fig:astrometric_accuracy}
\end{figure}

As shown in Fig.~\ref{fig:astrometric_accuracy}, the residual distributions show a slight elongation along the RA axis. This elongation is present at comparable levels in both the propagated and non-propagated solutions, confirming that it is unrelated to epoch propagation. \texttt{PhoPS} does not model or correct for this effect; it is therefore treated here as a telescope- or instrument-specific systematic effect outside the scope of the current pipeline.

\subsection{Photometric accuracy and zero-point magnitude}\label{subsec:PhotAccuracy}

We evaluated the photometric performance of \texttt{PhoPS} using the TUG100 dataset through both a population-level analysis and representative individual examples. The primary validation is based on the reference-star population retained for the photometric calibration, while the asteroid (201)~Penelope and the field star UCAC4 532-045441 are used as illustrative cases of the pipeline performance for a moving target and a stellar source, respectively. Additional test cases are presented in Appendix~\ref{sec:appendix}.

For the population-level uncertainty validation, residuals were computed for each retained reference star with respect to its weighted mean calibrated magnitude and were normalised by the total photometric uncertainty reported by \texttt{PhoPS}:
\begin{equation}
z_{ij} = \frac{m_{ij}-\bar{m}_i}{\sigma_{{\rm total},ij}},
\label{eq_zij}
\end{equation}
where $m_{ij}$ and $\sigma_{{\rm total},ij}$ are the calibrated magnitude and total uncertainty of star $i$ in frame $j$, and $\bar{m}_i$ is the weighted mean calibrated magnitude of that star over the sequence. For completeness, the full-sample normalised residual distribution of the final stability-filtered calibration-inlier sample has $\langle z\rangle = 0.014$, $\sigma_z = 1.106$, and $\mathrm{Var}(z)=1.222$. The mean value close to zero indicates that the calibrated magnitudes are unbiased on average. The aggregate width, however, combines magnitude intervals with different residual behaviour and therefore should not be interpreted as evidence for a globally well-scaled uncertainty model. The uncertainty model is instead evaluated primarily from the magnitude-binned distributions discussed below.

\begin{figure*}
\centering
\includegraphics[width=\linewidth]{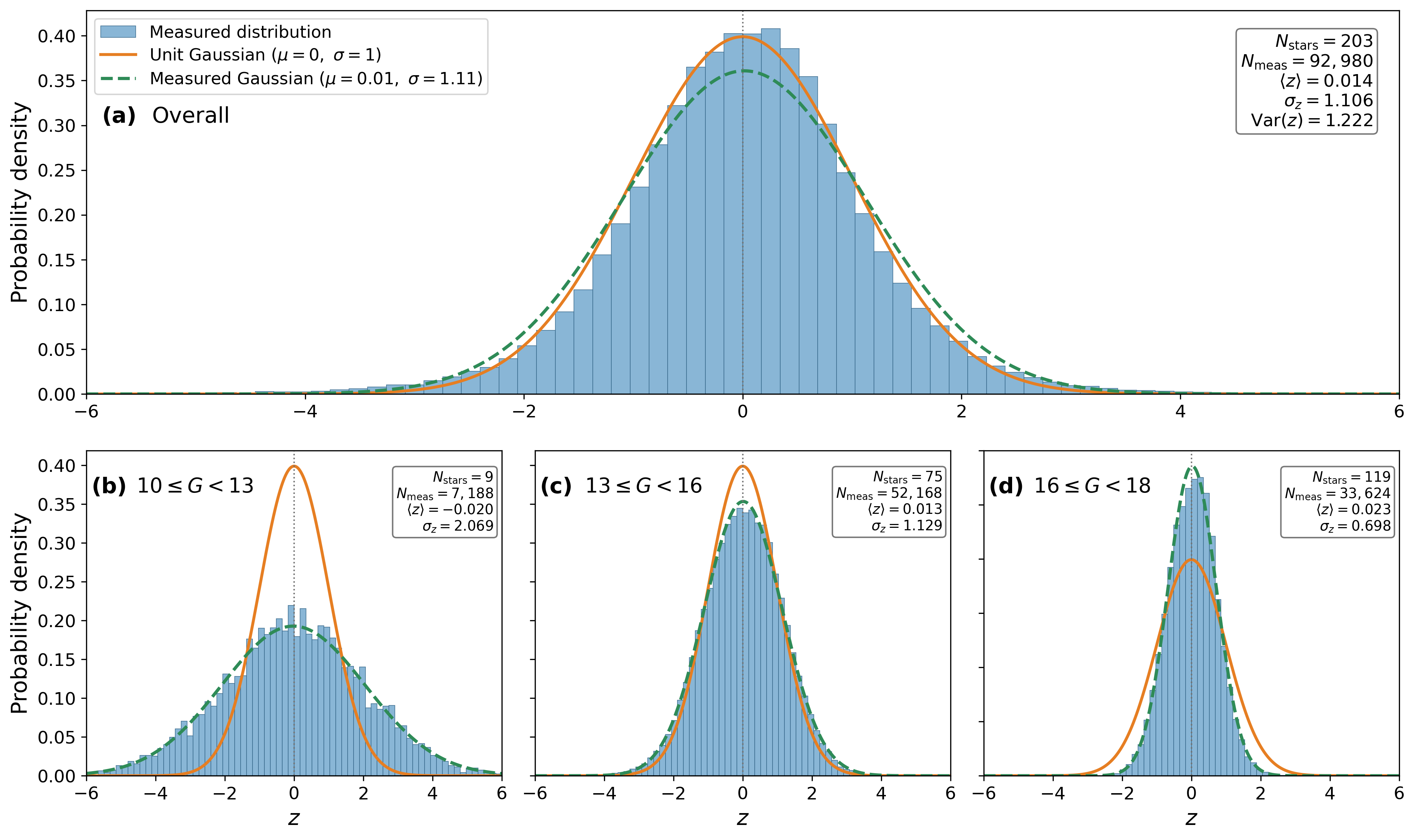}
\caption{Population-level validation of the total photometric uncertainties using the final stability-filtered calibration-inlier reference-star sample. Panel (a) shows the distribution of the normalised residuals for the full sample, where $z_{ij}=(m_{ij}-\bar{m}_i)/\sigma_{{\rm total},ij}$. The solid orange curve represents the unit Gaussian expected for a perfectly scaled uncertainty model, while the dashed green curve shows the Gaussian corresponding to the measured mean and standard deviation of the sample. Panels (b)--(d) show the same distribution separately for the three Gaia $G$-magnitude bins used in the analysis. The comparison reveals magnitude-dependent uncertainty behaviour: the intermediate-magnitude bin is closest to unit variance, the reported uncertainties are conservative for the faintest bin, and they underestimate the observed scatter for the limited bright-star bin.}

\label{fig:photometric_uncertainty_validation}
\end{figure*}

The residual distribution is magnitude dependent. The brightest bin, $10 \leq G < 13$, has $\sigma_z=2.069$, indicating that the reported uncertainties underestimate the observed scatter for the brightest reference stars. Because this bin contains only nine stars, the magnitude of the bright-end excess should be interpreted cautiously. The intermediate-magnitude bin, $13 \leq G < 16$, is closest to unit variance, with $\sigma_z=1.129$, although this result alone does not demonstrate that the uncertainty model is globally consistent. The faintest bin, $16 \leq G < 18$, is conservative, with $\sigma_z=0.698$, indicating that the reported uncertainties exceed the observed scatter in this magnitude range. Thus, the validation reveals a magnitude-dependent uncertainty behaviour rather than a single globally consistent error model. Applying a per-star three-sigma clipping changes the full-sample standard deviation only slightly, from 1.106 to 1.091, showing that the aggregate width is not driven by a small number of extreme measurements. The corresponding full-sample and magnitude-binned statistics are summarised in Table~\ref{tab:photometric_uncertainty_validation}, while the residual distributions are shown in Fig.~\ref{fig:photometric_uncertainty_validation}.

\begin{table}
\caption{Population-level validation of the photometric uncertainties.}
\label{tab:photometric_uncertainty_validation}
\scriptsize
\setlength{\tabcolsep}{7pt}

\begin{tabular}{lrrrrrrr}
\hline\hline
$G$ bin &
$N_\star$ &
$N_{\rm meas}$ &
$\langle z\rangle$ &
$\sigma_z$ &
$\mathrm{Var}(z)$ &
$\sigma_{z,3\sigma}$ &
$\mathrm{Var}(z)_{3\sigma}$ \\
\hline
Overall
& 203 & 92\,980 & 0.014 & 1.106 & 1.222 & 1.091 & 1.190 \\

$10$--$13$
& 9 & 7\,188 & -0.020 & 2.069 & 4.282 & 2.018 & 4.071 \\

$13$--$16$
& 75 & 52\,168 & 0.013 & 1.129 & 1.274 & 1.118 & 1.249 \\

$16$--$18$
& 119 & 33\,624 & 0.023 & 0.698 & 0.487 & 0.697 & 0.485 \\
\hline
\end{tabular}

\tablefoot{
The normalised residual is defined as $z_{ij}=(m_{ij}-\bar{m}_i)/\sigma_{{\rm total},ij}$. The columns headed by $3\sigma$ give the corresponding statistics after applying per-star $3\sigma$ clipping. The magnitude bins correspond to $10\leq G<13$, $13\leq G<16$, and $16\leq G<18$.}
\end{table}

\begin{figure*}
\centering
\includegraphics[width=\linewidth]{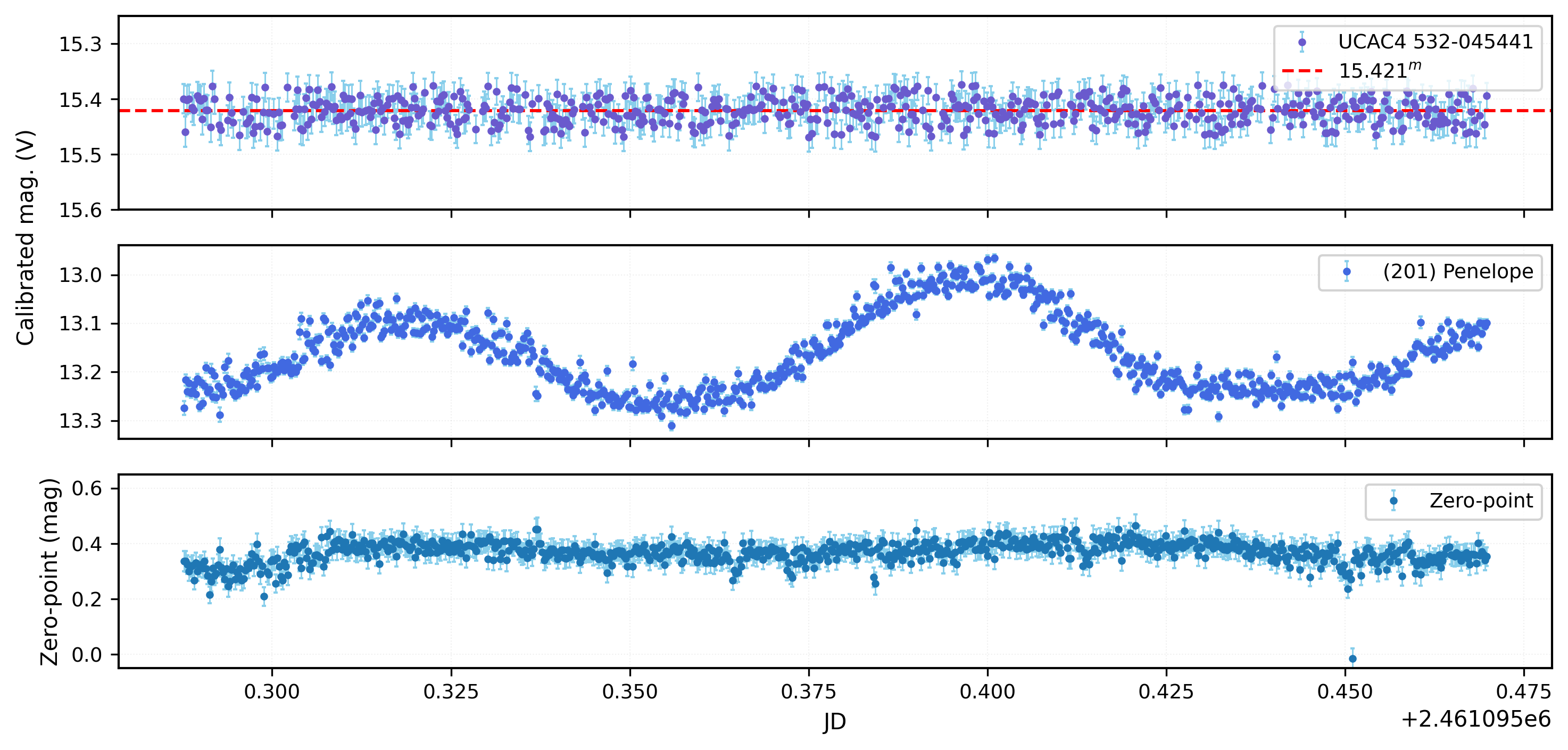}
\caption{Time-series photometric analysis of UCAC4 532-045441, (201)~Penelope, and the frame-by-frame photometric zero point. Top panel: calibrated light curve of UCAC4 532-045441; the dashed red line indicates its catalogued $V$-band magnitude. Middle panel: calibrated light curve of (201)~Penelope. Bottom panel: field-dependent photometric zero point evaluated at the asteroid position for each frame.}
\label{fig:penelope_LC}
\end{figure*}

The resulting calibrated light curves of (201)~Penelope and UCAC4 532-045441, together with the frame-by-frame variation of the photometric zero point, are shown in Fig.~\ref{fig:penelope_LC}. These examples are used only as representative checks of the pipeline behaviour for a moving target and a non-moving field source observed in the same frames; the quantitative uncertainty validation is given above using the full reference-star population.

For (201)~Penelope, \texttt{PhoPS} gives a mean calibrated magnitude of 13.165 mag over 837 frames. The calibrated time series spans a peak-to-peak range of 0.345 mag, while the mean total photometric uncertainty is 0.0091 mag. According to JPL Horizons, the predicted visual magnitude at the time of observation was $V = 13.315$ mag; the measured time series is of the expected order and clearly resolves the asteroid's variability.

For the field star UCAC4 532-045441, measured in the same observing sequence, \texttt{PhoPS} gives a mean calibrated magnitude of 15.421 mag over the 837 frames common to the Penelope analysis. This is in excellent agreement with its catalogued $V$-band magnitude of 15.421 mag. The mean total photometric uncertainty is 0.0250 mag. The absence of a coherent trend in the UCAC4 light curve provides a qualitative check that the variability observed for (201)~Penelope is not caused by frame-level photometric artefacts.

The zero point shown in the bottom panel is evaluated at the position of (201)~Penelope using the field-dependent radial model described in Sect.~\ref{sec:photocal}. Across the sequence, the mean zero point is 0.363 mag, with a standard deviation of 0.045 mag. Although moderate frame-to-frame zero-point variations are present, they do not introduce correlated features into the calibrated light curves. This indicates that \texttt{PhoPS} maintains a stable calibrated photometric scale in the presence of short-term degradations in observing conditions and that such zero-point fluctuations do not appear as coherent spurious features in the calibrated light curves.

\section{Discussion}\label{sec:discussion}

The \texttt{PhoPS} algorithm has already been applied in independent observational studies targeting both near-Earth and trans-Neptunian objects. In \cite{erece2023}, time-series photometry of the slow-rotating asteroid (2059)~Baboquivari was analysed using data from the TUG100 and the RTT150 telescopes, while \cite{kilic2026} applied \texttt{PhoPS} to observations of the trans-Neptunian object (28978)~Ixion obtained with the ESO New Technology Telescope (NTT) and the ESO VLT UT2 telescope, deriving its absolute magnitude and broadband colours. Together with the validation results presented in this work, these applications demonstrate that the calibration strategy implemented in \texttt{PhoPS} provides consistent results across datasets obtained with different telescopes, detectors, and observing conditions. A key element of this approach is the use of the Gaia DR3 catalogue as a unified reference for both astrometric and photometric calibration, enabling a homogeneous reduction procedure without requiring manual reference star selection or the combination of multiple catalogues with different photometric systems. This unified calibration scheme is particularly advantageous for large time-series datasets, where small systematic inconsistencies between reduction steps can otherwise propagate into the final scientific products. The results obtained for both stellar and moving targets indicate that the adopted strategy provides reliable calibration while preserving the scientific integrity of the resulting light curves.

Another key feature of \texttt{PhoPS} is the dynamic generation of local Gaia DR3-based astrometric index files propagated to the epoch of observation. This strategy provides an epoch-consistent reference for the Astrometry.net workflow, which is particularly useful for multi-year datasets and fields containing stars with non-negligible proper motions. In addition, the pipeline remains portable because it does not require a permanent pre-installed astrometric index collection. When multiple images of the same field are processed, these locally generated index files are identified through coordinate-based flags and reused, improving the efficiency of sequential reductions. In parallel, \texttt{PhoPS} models spatial variations in the photometric zero point across the detector plane. The RANSAC-based radial model provides a robust and computationally efficient way to characterise large-scale systematics, such as vignetting and detector non-uniformities, using the distribution of matched reference stars. Validation tests show that this approach supports a stable photometric solution and helps mitigate the impact of instrumental effects and short-term atmospheric variations. The population-level validation also shows that photometric calibration accuracy and uncertainty calibration should be considered separately. Although the normalised residuals are centred close to zero, indicating unbiased calibrated magnitudes on average, their width depends on source brightness. The reported uncertainties underestimate the observed scatter for the limited bright-star sample and are conservative for the faintest reference stars, while the intermediate-magnitude bin is closest to unit variance. Consequently, the full-sample value of $\mathrm{Var}(z)=1.222$ should not be interpreted as evidence for a single globally consistent uncertainty model.

In addition to its scientific reduction capabilities, \texttt{PhoPS} can also serve as a diagnostic tool for telescope systems. The pipeline outputs, including astrometric residuals and spatially varying photometric zero-point models, provide quantitative information about the optical and mechanical performance of the instrument. Systematic patterns in these quantities can reveal issues related to collimation, tracking, focus stability, or vignetting. The modular design of the pipeline further allows it to be adapted for targeted verification tasks, making it suitable for routine quality control.

Despite these advantages, the current implementation of \texttt{PhoPS} has several limitations. When observations are obtained with filters lacking well-established transformations to standard photometric systems, the pipeline cannot provide calibrated magnitudes and instead defaults to differential photometry. This also applies to unfiltered (clear) observations. Furthermore, the transformation from Gaia photometry to standard passbands relies on empirical colour relations, which may introduce systematic uncertainties for sources with unusual spectral energy distributions or in regions affected by strong and spatially variable extinction. In such cases, these effects may manifest as outliers in the calibration process, which are mitigated through the RANSAC-based fitting procedure, although residual systematics may still persist under extreme conditions. The current field-dependent calibration also assumes that the spatial variation of the zero point can be represented by a linear function of radial distance from the detector centre. This radial model is intended as a first-order description for the moderate fields of view considered in this work, but it may not capture asymmetric or higher-order two-dimensional structures in very wide-field instruments, residual flat-field patterns, spatially variable transparency, or severely degraded observing conditions. Such datasets may require a higher-order or fully two-dimensional zero-point model supported by a sufficiently dense reference-star distribution.

\section{Conclusions}\label{sec:conclusion}

In this work, we presented \texttt{PhoPS}, an automated pipeline for photometric reduction with integrated astrometric calibration of astronomical imaging data. The pipeline combines Gaia DR3-based reference extraction, dynamic astrometric index generation, and a field-dependent photometric zero-point model within a unified reduction framework. This design enables homogeneous processing without requiring manually selected reference stars or a permanent pre-installed astrometric index collection.

Validation tests performed on data from different telescope systems show that \texttt{PhoPS} provides consistent astrometric solutions and calibrated photometric measurements under a variety of observing conditions, while also revealing magnitude-dependent limitations in the reported photometric uncertainties. In particular, the dynamic indexing strategy provides epoch-consistent local Gaia DR3-based astrometric references and improves portability, while the RANSAC-based radial zero-point model accounts for large-scale spatial systematics across the detector plane. The photometric tests further show that, in the examined tracking-error case, short-term image degradation did not propagate into the final calibrated light curves. Together with its successful application in previous observational studies, these results indicate that \texttt{PhoPS} is well suited to the routine reduction of both stellar and moving target observations, especially in time-series datasets where homogeneous calibration and resistance to systematic effects are essential. Future developments may include support for additional photometric systems and optional higher-order astrometric refinement, further extending the applicability of the pipeline in the survey era.

\begin{acknowledgements}
We express our gratitude to the Türkiye National Observatories and its staff for the valuable contributions provided through the infrastructure usage and additional data sharing support offered at the TUG (TÜBİTAK National Observatory, Antalya) site in the realisation of this study. Y.K. acknowledges financial support from the Severo Ochoa grant CEX2021-001131-S (MCIN/AEI/10.13039/501100011033). Y.K. also acknowledges support from the Spanish I+D+i project PID2022-139555NB-I00 (TNO-JWST) funded by MCIN/AEI/10.13039/501100011033. This research has made use of data from the European Space Agency (ESA) mission Gaia\footnote{\url{https://www.cosmos.esa.int/gaia}}, processed by the Gaia Data Processing and Analysis Consortium\footnote{\url{https://www.cosmos.esa.int/web/gaia/dpac/consortium}}, with funding provided by institutions participating in the Gaia Multilateral Agreement. This work is based in part on observations collected at the European Southern Observatory under ESO programmes 075.C-0431(A) and 178.C-0036(N), including data obtained from the ESO Science Archive Facility. We also thank the anonymous referee for a careful and constructive review that helped improve the clarity and robustness of the manuscript.
\end{acknowledgements}

\bibliographystyle{bibtex/aa}
\bibliography{bibtex/references}

@article{assafin2023,
title = {Astrometry with PRAIA},
journal = {Planetary and Space Science},
volume = {238},
pages = {105801},
year = {2023},
issn = {0032-0633},
doi = {https://doi.org/10.1016/j.pss.2023.105801},
url = {https://www.sciencedirect.com/science/article/pii/S0032063323001708},
author = {M. Assafin}
}

@ARTICLE{Gaia2018,
       author = {{Gaia Collaboration} and {Brown}, A.~G.~A. and {Vallenari}, A. and {Prusti}, T. and {de Bruijne}, J.~H.~J. and {Babusiaux}, C. and {Bailer-Jones}, C.~A.~L. and {Biermann}, M. and {Evans}, D.~W. and {Eyer}, L. and {Jansen}, F. and {Jordi}, C. and {Klioner}, S.~A. and {Lammers}, U. and {Lindegren}, L. and {Luri}, X. and {Mignard}, F. and {Panem}, C. and {Pourbaix}, D. and {Randich}, S. and {Sartoretti}, P. and {Siddiqui}, H.~I. and {Soubiran}, C. and {van Leeuwen}, F. and {Walton}, N.~A. and {Arenou}, F. and {Bastian}, U. and {Cropper}, M. and {Drimmel}, R. and {Katz}, D. and {Lattanzi}, M.~G. and {Bakker}, J. and {Cacciari}, C. and {Casta{\~n}eda}, J. and {Chaoul}, L. and {Cheek}, N. and {De Angeli}, F. and {Fabricius}, C. and {Guerra}, R. and {Holl}, B. and {Masana}, E. and {Messineo}, R. and {Mowlavi}, N. and {Nienartowicz}, K. and {Panuzzo}, P. and {Portell}, J. and {Riello}, M. and {Seabroke}, G.~M. and {Tanga}, P. and {Th{\'e}venin}, F. and {Gracia-Abril}, G. and {Comoretto}, G. and {Garcia-Reinaldos}, M. and {Teyssier}, D. and {Altmann}, M. and {Andrae}, R. and {Audard}, M. and {Bellas-Velidis}, I. and {Benson}, K. and {Berthier}, J. and {Blomme}, R. and {Burgess}, P. and {Busso}, G. and {Carry}, B. and {Cellino}, A. and {Clementini}, G. and {Clotet}, M. and {Creevey}, O. and {Davidson}, M. and {De Ridder}, J. and {Delchambre}, L. and {Dell'Oro}, A. and {Ducourant}, C. and {Fern{\'a}ndez-Hern{\'a}ndez}, J. and {Fouesneau}, M. and {Fr{\'e}mat}, Y. and {Galluccio}, L. and {Garc{\'\i}a-Torres}, M. and {Gonz{\'a}lez-N{\'u}{\~n}ez}, J. and {Gonz{\'a}lez-Vidal}, J.~J. and {Gosset}, E. and {Guy}, L.~P. and {Halbwachs}, J.-L. and {Hambly}, N.~C. and {Harrison}, D.~L. and {Hern{\'a}ndez}, J. and {Hestroffer}, D. and {Hodgkin}, S.~T. and {Hutton}, A. and {Jasniewicz}, G. and {Jean-Antoine-Piccolo}, A. and {Jordan}, S. and {Korn}, A.~J. and {Krone-Martins}, A. and {Lanzafame}, A.~C. and {Lebzelter}, T. and {L{\"o}ffler}, W. and {Manteiga}, M. and {Marrese}, P.~M. and {Mart{\'\i}n-Fleitas}, J.~M. and {Moitinho}, A. and {Mora}, A. and {Muinonen}, K. and {Osinde}, J. and {Pancino}, E. and {Pauwels}, T. and {Petit}, J.-M. and {Recio-Blanco}, A. and {Richards}, P.~J. and {Rimoldini}, L. and {Robin}, A.~C. and {Sarro}, L.~M. and {Siopis}, C. and {Smith}, M. and {Sozzetti}, A. and {S{\"u}veges}, M. and {Torra}, J. and {van Reeven}, W. and {Abbas}, U. and {Abreu Aramburu}, A. and {Accart}, S. and {Aerts}, C. and {Altavilla}, G. and {{\'A}lvarez}, M.~A. and {Alvarez}, R. and {Alves}, J. and {Anderson}, R.~I. and {Andrei}, A.~H. and {Anglada Varela}, E. and {Antiche}, E. and {Antoja}, T. and {Arcay}, B. and {Astraatmadja}, T.~L. and {Bach}, N. and {Baker}, S.~G. and {Balaguer-N{\'u}{\~n}ez}, L. and {Balm}, P. and {Barache}, C. and {Barata}, C. and {Barbato}, D. and {Barblan}, F. and {Barklem}, P.~S. and {Barrado}, D. and {Barros}, M. and {Barstow}, M.~A. and {Bartholom{\'e} Mu{\~n}oz}, S. and {Bassilana}, J.-L. and {Becciani}, U. and {Bellazzini}, M. and {Berihuete}, A. and {Bertone}, S. and {Bianchi}, L. and {Bienaym{\'e}}, O. and {Blanco-Cuaresma}, S. and {Boch}, T. and {Boeche}, C. and {Bombrun}, A. and {Borrachero}, R. and {Bossini}, D. and {Bouquillon}, S. and {Bourda}, G. and {Bragaglia}, A. and {Bramante}, L. and {Breddels}, M.~A. and {Bressan}, A. and {Brouillet}, N. and {Br{\"u}semeister}, T. and {Brugaletta}, E. and {Bucciarelli}, B. and {Burlacu}, A. and {Busonero}, D. and {Butkevich}, A.~G. and {Buzzi}, R. and {Caffau}, E. and {Cancelliere}, R. and {Cannizzaro}, G. and {Cantat-Gaudin}, T. and {Carballo}, R. and {Carlucci}, T. and {Carrasco}, J.~M. and {Casamiquela}, L. and {Castellani}, M. and {Castro-Ginard}, A. and {Charlot}, P. and {Chemin}, L. and {Chiavassa}, A. and {Cocozza}, G. and {Costigan}, G. and {Cowell}, S. and {Crifo}, F. and {Crosta}, M. and {Crowley}, C. and {Cuypers}, J. and {Dafonte}, C. and {Damerdji}, Y. and {Dapergolas}, A. and {David}, P. and {David}, M. and {de Laverny}, P. and {De Luise}, F.},
        title = "{Gaia Data Release 2. Summary of the contents and survey properties}",
      journal = {\aap},
         year = 2018,
        month = aug,
       volume = {616},
          eid = {A1},
        pages = {A1},
          doi = {10.1051/0004-6361/201833051},
archivePrefix = {arXiv},
       eprint = {1804.09365},
 primaryClass = {astro-ph.GA},
       adsurl = {https://ui.adsabs.harvard.edu/abs/2018A&A...616A...1G}
}

@ARTICLE{Gaia2016b,
       author = {{Gaia Collaboration} and {Brown}, A.~G.~A. and {Vallenari}, A. and {Prusti}, T. and {de Bruijne}, J.~H.~J. and {Mignard}, F. and {Drimmel}, R. and {Babusiaux}, C. and {Bailer-Jones}, C.~A.~L. and {Bastian}, U. and {Biermann}, M. and {Evans}, D.~W. and {Eyer}, L. and {Jansen}, F. and {Jordi}, C. and {Katz}, D. and {Klioner}, S.~A. and {Lammers}, U. and {Lindegren}, L. and {Luri}, X. and {O'Mullane}, W. and {Panem}, C. and {Pourbaix}, D. and {Randich}, S. and {Sartoretti}, P. and {Siddiqui}, H.~I. and {Soubiran}, C. and {Valette}, V. and {van Leeuwen}, F. and {Walton}, N.~A. and {Aerts}, C. and {Arenou}, F. and {Cropper}, M. and {H{\o}g}, E. and {Lattanzi}, M.~G. and {Grebel}, E.~K. and {Holland}, A.~D. and {Huc}, C. and {Passot}, X. and {Perryman}, M. and {Bramante}, L. and {Cacciari}, C. and {Casta{\~n}eda}, J. and {Chaoul}, L. and {Cheek}, N. and {De Angeli}, F. and {Fabricius}, C. and {Guerra}, R. and {Hern{\'a}ndez}, J. and {Jean-Antoine-Piccolo}, A. and {Masana}, E. and {Messineo}, R. and {Mowlavi}, N. and {Nienartowicz}, K. and {Ord{\'o}{\~n}ez-Blanco}, D. and {Panuzzo}, P. and {Portell}, J. and {Richards}, P.~J. and {Riello}, M. and {Seabroke}, G.~M. and {Tanga}, P. and {Th{\'e}venin}, F. and {Torra}, J. and {Els}, S.~G. and {Gracia-Abril}, G. and {Comoretto}, G. and {Garcia-Reinaldos}, M. and {Lock}, T. and {Mercier}, E. and {Altmann}, M. and {Andrae}, R. and {Astraatmadja}, T.~L. and {Bellas-Velidis}, I. and {Benson}, K. and {Berthier}, J. and {Blomme}, R. and {Busso}, G. and {Carry}, B. and {Cellino}, A. and {Clementini}, G. and {Cowell}, S. and {Creevey}, O. and {Cuypers}, J. and {Davidson}, M. and {De Ridder}, J. and {de Torres}, A. and {Delchambre}, L. and {Dell'Oro}, A. and {Ducourant}, C. and {Fr{\'e}mat}, Y. and {Garc{\'\i}a-Torres}, M. and {Gosset}, E. and {Halbwachs}, J.-L. and {Hambly}, N.~C. and {Harrison}, D.~L. and {Hauser}, M. and {Hestroffer}, D. and {Hodgkin}, S.~T. and {Huckle}, H.~E. and {Hutton}, A. and {Jasniewicz}, G. and {Jordan}, S. and {Kontizas}, M. and {Korn}, A.~J. and {Lanzafame}, A.~C. and {Manteiga}, M. and {Moitinho}, A. and {Muinonen}, K. and {Osinde}, J. and {Pancino}, E. and {Pauwels}, T. and {Petit}, J.-M. and {Recio-Blanco}, A. and {Robin}, A.~C. and {Sarro}, L.~M. and {Siopis}, C. and {Smith}, M. and {Smith}, K.~W. and {Sozzetti}, A. and {Thuillot}, W. and {van Reeven}, W. and {Viala}, Y. and {Abbas}, U. and {Abreu Aramburu}, A. and {Accart}, S. and {Aguado}, J.~J. and {Allan}, P.~M. and {Allasia}, W. and {Altavilla}, G. and {{\'A}lvarez}, M.~A. and {Alves}, J. and {Anderson}, R.~I. and {Andrei}, A.~H. and {Anglada Varela}, E. and {Antiche}, E. and {Antoja}, T. and {Ant{\'o}n}, S. and {Arcay}, B. and {Bach}, N. and {Baker}, S.~G. and {Balaguer-N{\'u}{\~n}ez}, L. and {Barache}, C. and {Barata}, C. and {Barbier}, A. and {Barblan}, F. and {Barrado y Navascu{\'e}s}, D. and {Barros}, M. and {Barstow}, M.~A. and {Becciani}, U. and {Bellazzini}, M. and {Bello Garc{\'\i}a}, A. and {Belokurov}, V. and {Bendjoya}, P. and {Berihuete}, A. and {Bianchi}, L. and {Bienaym{\'e}}, O. and {Billebaud}, F. and {Blagorodnova}, N. and {Blanco-Cuaresma}, S. and {Boch}, T. and {Bombrun}, A. and {Borrachero}, R. and {Bouquillon}, S. and {Bourda}, G. and {Bouy}, H. and {Bragaglia}, A. and {Breddels}, M.~A. and {Brouillet}, N. and {Br{\"u}semeister}, T. and {Bucciarelli}, B. and {Burgess}, P. and {Burgon}, R. and {Burlacu}, A. and {Busonero}, D. and {Buzzi}, R. and {Caffau}, E. and {Cambras}, J. and {Campbell}, H. and {Cancelliere}, R. and {Cantat-Gaudin}, T. and {Carlucci}, T. and {Carrasco}, J.~M. and {Castellani}, M. and {Charlot}, P. and {Charnas}, J. and {Chiavassa}, A. and {Clotet}, M. and {Cocozza}, G. and {Collins}, R.~S. and {Costigan}, G. and {Crifo}, F. and {Cross}, N.~J.~G. and {Crosta}, M. and {Crowley}, C. and {Dafonte}, C. and {Damerdji}, Y. and {Dapergolas}, A. and {David}, P. and {David}, M. and {De Cat}, P.},
        title = "{Gaia Data Release 1. Summary of the astrometric, photometric, and survey properties}",
      journal = {\aap},
         year = 2016,
        month = nov,
       volume = {595},
          eid = {A2},
        pages = {A2},
          doi = {10.1051/0004-6361/201629512},
archivePrefix = {arXiv},
       eprint = {1609.04172},
 primaryClass = {astro-ph.IM},
       adsurl = {https://ui.adsabs.harvard.edu/abs/2016A&A...595A...2G}
}

@ARTICLE{Gaia2016a,
       author = {{Gaia Collaboration} and {Prusti}, T. and {de Bruijne}, J.~H.~J. and {Brown}, A.~G.~A. and {Vallenari}, A. and {Babusiaux}, C. and {Bailer-Jones}, C.~A.~L. and {Bastian}, U. and {Biermann}, M. and {Evans}, D.~W. and {Eyer}, L. and {Jansen}, F. and {Jordi}, C. and {Klioner}, S.~A. and {Lammers}, U. and {Lindegren}, L. and {Luri}, X. and {Mignard}, F. and {Milligan}, D.~J. and {Panem}, C. and {Poinsignon}, V. and {Pourbaix}, D. and {Randich}, S. and {Sarri}, G. and {Sartoretti}, P. and {Siddiqui}, H.~I. and {Soubiran}, C. and {Valette}, V. and {van Leeuwen}, F. and {Walton}, N.~A. and {Aerts}, C. and {Arenou}, F. and {Cropper}, M. and {Drimmel}, R. and {H{\o}g}, E. and {Katz}, D. and {Lattanzi}, M.~G. and {O'Mullane}, W. and {Grebel}, E.~K. and {Holland}, A.~D. and {Huc}, C. and {Passot}, X. and {Bramante}, L. and {Cacciari}, C. and {Casta{\~n}eda}, J. and {Chaoul}, L. and {Cheek}, N. and {De Angeli}, F. and {Fabricius}, C. and {Guerra}, R. and {Hern{\'a}ndez}, J. and {Jean-Antoine-Piccolo}, A. and {Masana}, E. and {Messineo}, R. and {Mowlavi}, N. and {Nienartowicz}, K. and {Ord{\'o}{\~n}ez-Blanco}, D. and {Panuzzo}, P. and {Portell}, J. and {Richards}, P.~J. and {Riello}, M. and {Seabroke}, G.~M. and {Tanga}, P. and {Th{\'e}venin}, F. and {Torra}, J. and {Els}, S.~G. and {Gracia-Abril}, G. and {Comoretto}, G. and {Garcia-Reinaldos}, M. and {Lock}, T. and {Mercier}, E. and {Altmann}, M. and {Andrae}, R. and {Astraatmadja}, T.~L. and {Bellas-Velidis}, I. and {Benson}, K. and {Berthier}, J. and {Blomme}, R. and {Busso}, G. and {Carry}, B. and {Cellino}, A. and {Clementini}, G. and {Cowell}, S. and {Creevey}, O. and {Cuypers}, J. and {Davidson}, M. and {De Ridder}, J. and {de Torres}, A. and {Delchambre}, L. and {Dell'Oro}, A. and {Ducourant}, C. and {Fr{\'e}mat}, Y. and {Garc{\'\i}a-Torres}, M. and {Gosset}, E. and {Halbwachs}, J.-L. and {Hambly}, N.~C. and {Harrison}, D.~L. and {Hauser}, M. and {Hestroffer}, D. and {Hodgkin}, S.~T. and {Huckle}, H.~E. and {Hutton}, A. and {Jasniewicz}, G. and {Jordan}, S. and {Kontizas}, M. and {Korn}, A.~J. and {Lanzafame}, A.~C. and {Manteiga}, M. and {Moitinho}, A. and {Muinonen}, K. and {Osinde}, J. and {Pancino}, E. and {Pauwels}, T. and {Petit}, J.-M. and {Recio-Blanco}, A. and {Robin}, A.~C. and {Sarro}, L.~M. and {Siopis}, C. and {Smith}, M. and {Smith}, K.~W. and {Sozzetti}, A. and {Thuillot}, W. and {van Reeven}, W. and {Viala}, Y. and {Abbas}, U. and {Abreu Aramburu}, A. and {Accart}, S. and {Aguado}, J.~J. and {Allan}, P.~M. and {Allasia}, W. and {Altavilla}, G. and {{\'A}lvarez}, M.~A. and {Alves}, J. and {Anderson}, R.~I. and {Andrei}, A.~H. and {Anglada Varela}, E. and {Antiche}, E. and {Antoja}, T. and {Ant{\'o}n}, S. and {Arcay}, B. and {Atzei}, A. and {Ayache}, L. and {Bach}, N. and {Baker}, S.~G. and {Balaguer-N{\'u}{\~n}ez}, L. and {Barache}, C. and {Barata}, C. and {Barbier}, A. and {Barblan}, F. and {Baroni}, M. and {Barrado y Navascu{\'e}s}, D. and {Barros}, M. and {Barstow}, M.~A. and {Becciani}, U. and {Bellazzini}, M. and {Bellei}, G. and {Bello Garc{\'\i}a}, A. and {Belokurov}, V. and {Bendjoya}, P. and {Berihuete}, A. and {Bianchi}, L. and {Bienaym{\'e}}, O. and {Billebaud}, F. and {Blagorodnova}, N. and {Blanco-Cuaresma}, S. and {Boch}, T. and {Bombrun}, A. and {Borrachero}, R. and {Bouquillon}, S. and {Bourda}, G. and {Bouy}, H. and {Bragaglia}, A. and {Breddels}, M.~A. and {Brouillet}, N. and {Br{\"u}semeister}, T. and {Bucciarelli}, B. and {Budnik}, F. and {Burgess}, P. and {Burgon}, R. and {Burlacu}, A. and {Busonero}, D. and {Buzzi}, R. and {Caffau}, E. and {Cambras}, J. and {Campbell}, H. and {Cancelliere}, R. and {Cantat-Gaudin}, T. and {Carlucci}, T. and {Carrasco}, J.~M. and {Castellani}, M. and {Charlot}, P. and {Charnas}, J. and {Charvet}, P. and {Chassat}, F. and {Chiavassa}, A. and {Clotet}, M. and {Cocozza}, G. and {Collins}, R.~S. and {Collins}, P. and {Costigan}, G.},
        title = "{The Gaia mission}",
      journal = {\aap},
         year = 2016,
        month = nov,
       volume = {595},
          eid = {A1},
        pages = {A1},
          doi = {10.1051/0004-6361/201629272},
archivePrefix = {arXiv},
       eprint = {1609.04153},
 primaryClass = {astro-ph.IM},
       adsurl = {https://ui.adsabs.harvard.edu/abs/2016A&A...595A...1G}
}

@ARTICLE{Gaia2023,
       author = {{Gaia Collaboration} and {Vallenari}, A. and {Brown}, A.~G.~A. and {Prusti}, T. and {de Bruijne}, J.~H.~J. and {Arenou}, F. and {Babusiaux}, C. and {Biermann}, M. and {Creevey}, O.~L. and {Ducourant}, C. and {Evans}, D.~W. and {Eyer}, L. and {Guerra}, R. and {Hutton}, A. and {Jordi}, C. and {Klioner}, S.~A. and {Lammers}, U.~L. and {Lindegren}, L. and {Luri}, X. and {Mignard}, F. and {Panem}, C. and {Pourbaix}, D. and {Randich}, S. and {Sartoretti}, P. and {Soubiran}, C. and {Tanga}, P. and {Walton}, N.~A. and {Bailer-Jones}, C.~A.~L. and {Bastian}, U. and {Drimmel}, R. and {Jansen}, F. and {Katz}, D. and {Lattanzi}, M.~G. and {van Leeuwen}, F. and {Bakker}, J. and {Cacciari}, C. and {Casta{\~n}eda}, J. and {De Angeli}, F. and {Fabricius}, C. and {Fouesneau}, M. and {Fr{\'e}mat}, Y. and {Galluccio}, L. and {Guerrier}, A. and {Heiter}, U. and {Masana}, E. and {Messineo}, R. and {Mowlavi}, N. and {Nicolas}, C. and {Nienartowicz}, K. and {Pailler}, F. and {Panuzzo}, P. and {Riclet}, F. and {Roux}, W. and {Seabroke}, G.~M. and {Sordo}, R. and {Th{\'e}venin}, F. and {Gracia-Abril}, G. and {Portell}, J. and {Teyssier}, D. and {Altmann}, M. and {Andrae}, R. and {Audard}, M. and {Bellas-Velidis}, I. and {Benson}, K. and {Berthier}, J. and {Blomme}, R. and {Burgess}, P.~W. and {Busonero}, D. and {Busso}, G. and {C{\'a}novas}, H. and {Carry}, B. and {Cellino}, A. and {Cheek}, N. and {Clementini}, G. and {Damerdji}, Y. and {Davidson}, M. and {de Teodoro}, P. and {Nu{\~n}ez Campos}, M. and {Delchambre}, L. and {Dell'Oro}, A. and {Esquej}, P. and {Fern{\'a}ndez-Hern{\'a}ndez}, J. and {Fraile}, E. and {Garabato}, D. and {Garc{\'\i}a-Lario}, P. and {Gosset}, E. and {Haigron}, R. and {Halbwachs}, J.-L. and {Hambly}, N.~C. and {Harrison}, D.~L. and {Hern{\'a}ndez}, J. and {Hestroffer}, D. and {Hodgkin}, S.~T. and {Holl}, B. and {Jan{\ss}en}, K. and {Jevardat de Fombelle}, G. and {Jordan}, S. and {Krone-Martins}, A. and {Lanzafame}, A.~C. and {L{\"o}ffler}, W. and {Marchal}, O. and {Marrese}, P.~M. and {Moitinho}, A. and {Muinonen}, K. and {Osborne}, P. and {Pancino}, E. and {Pauwels}, T. and {Recio-Blanco}, A. and {Reyl{\'e}}, C. and {Riello}, M. and {Rimoldini}, L. and {Roegiers}, T. and {Rybizki}, J. and {Sarro}, L.~M. and {Siopis}, C. and {Smith}, M. and {Sozzetti}, A. and {Utrilla}, E. and {van Leeuwen}, M. and {Abbas}, U. and {{\'A}brah{\'a}m}, P. and {Abreu Aramburu}, A. and {Aerts}, C. and {Aguado}, J.~J. and {Ajaj}, M. and {Aldea-Montero}, F. and {Altavilla}, G. and {{\'A}lvarez}, M.~A. and {Alves}, J. and {Anders}, F. and {Anderson}, R.~I. and {Anglada Varela}, E. and {Antoja}, T. and {Baines}, D. and {Baker}, S.~G. and {Balaguer-N{\'u}{\~n}ez}, L. and {Balbinot}, E. and {Balog}, Z. and {Barache}, C. and {Barbato}, D. and {Barros}, M. and {Barstow}, M.~A. and {Bartolom{\'e}}, S. and {Bassilana}, J.-L. and {Bauchet}, N. and {Becciani}, U. and {Bellazzini}, M. and {Berihuete}, A. and {Bernet}, M. and {Bertone}, S. and {Bianchi}, L. and {Binnenfeld}, A. and {Blanco-Cuaresma}, S. and {Blazere}, A. and {Boch}, T. and {Bombrun}, A. and {Bossini}, D. and {Bouquillon}, S. and {Bragaglia}, A. and {Bramante}, L. and {Breedt}, E. and {Bressan}, A. and {Brouillet}, N. and {Brugaletta}, E. and {Bucciarelli}, B. and {Burlacu}, A. and {Butkevich}, A.~G. and {Buzzi}, R. and {Caffau}, E. and {Cancelliere}, R. and {Cantat-Gaudin}, T. and {Carballo}, R. and {Carlucci}, T. and {Carnerero}, M.~I. and {Carrasco}, J.~M. and {Casamiquela}, L. and {Castellani}, M. and {Castro-Ginard}, A. and {Chaoul}, L. and {Charlot}, P. and {Chemin}, L. and {Chiaramida}, V. and {Chiavassa}, A. and {Chornay}, N. and {Comoretto}, G. and {Contursi}, G. and {Cooper}, W.~J. and {Cornez}, T. and {Cowell}, S. and {Crifo}, F. and {Cropper}, M. and {Crosta}, M. and {Crowley}, C. and {Dafonte}, C. and {Dapergolas}, A. and {David}, M. and {David}, P. and {de Laverny}, P. and {De Luise}, F. and {De March}, R.},
        title = "{Gaia Data Release 3. Summary of the content and survey properties}",
      journal = {\aap},
         year = 2023,
        month = jun,
       volume = {674},
          eid = {A1},
        pages = {A1},
          doi = {10.1051/0004-6361/202243940},
archivePrefix = {arXiv},
       eprint = {2208.00211},
 primaryClass = {astro-ph.GA},
       adsurl = {https://ui.adsabs.harvard.edu/abs/2023A&A...674A...1G}
}

@ARTICLE{2mass,
       author = {{Skrutskie}, M.~F. and {Cutri}, R.~M. and {Stiening}, R. and {Weinberg}, M.~D. and {Schneider}, S. and {Carpenter}, J.~M. and {Beichman}, C. and {Capps}, R. and {Chester}, T. and {Elias}, J. and {Huchra}, J. and {Liebert}, J. and {Lonsdale}, C. and {Monet}, D.~G. and {Price}, S. and {Seitzer}, P. and {Jarrett}, T. and {Kirkpatrick}, J.~D. and {Gizis}, J.~E. and {Howard}, E. and {Evans}, T. and {Fowler}, J. and {Fullmer}, L. and {Hurt}, R. and {Light}, R. and {Kopan}, E.~L. and {Marsh}, K.~A. and {McCallon}, H.~L. and {Tam}, R. and {Van Dyk}, S. and {Wheelock}, S.},
        title = "{The Two Micron All Sky Survey (2MASS)}",
      journal = {\aj},
         year = 2006,
        month = feb,
       volume = {131},
       number = {2},
        pages = {1163-1183},
          doi = {10.1086/498708},
       adsurl = {https://ui.adsabs.harvard.edu/abs/2006AJ....131.1163S}
}

@ARTICLE{sdss,
       author = {{York}, Donald G. and {Adelman}, J. and {Anderson}, Jr., John E. and {Anderson}, Scott F. and {Annis}, James and {Bahcall}, Neta A. and {Bakken}, J.~A. and {Barkhouser}, Robert and {Bastian}, Steven and {Berman}, Eileen and {Boroski}, William N. and {Bracker}, Steve and {Briegel}, Charlie and {Briggs}, John W. and {Brinkmann}, J. and {Brunner}, Robert and {Burles}, Scott and {Carey}, Larry and {Carr}, Michael A. and {Castander}, Francisco J. and {Chen}, Bing and {Colestock}, Patrick L. and {Connolly}, A.~J. and {Crocker}, J.~H. and {Csabai}, Istv{\'a}n and {Czarapata}, Paul C. and {Davis}, John Eric and {Doi}, Mamoru and {Dombeck}, Tom and {Eisenstein}, Daniel and {Ellman}, Nancy and {Elms}, Brian R. and {Evans}, Michael L. and {Fan}, Xiaohui and {Federwitz}, Glenn R. and {Fiscelli}, Larry and {Friedman}, Scott and {Frieman}, Joshua A. and {Fukugita}, Masataka and {Gillespie}, Bruce and {Gunn}, James E. and {Gurbani}, Vijay K. and {de Haas}, Ernst and {Haldeman}, Merle and {Harris}, Frederick H. and {Hayes}, J. and {Heckman}, Timothy M. and {Hennessy}, G.~S. and {Hindsley}, Robert B. and {Holm}, Scott and {Holmgren}, Donald J. and {Huang}, Chi-hao and {Hull}, Charles and {Husby}, Don and {Ichikawa}, Shin-Ichi and {Ichikawa}, Takashi and {Ivezi{\'c}}, {\v{Z}}eljko and {Kent}, Stephen and {Kim}, Rita S.~J. and {Kinney}, E. and {Klaene}, Mark and {Kleinman}, A.~N. and {Kleinman}, S. and {Knapp}, G.~R. and {Korienek}, John and {Kron}, Richard G. and {Kunszt}, Peter Z. and {Lamb}, D.~Q. and {Lee}, B. and {Leger}, R. French and {Limmongkol}, Siriluk and {Lindenmeyer}, Carl and {Long}, Daniel C. and {Loomis}, Craig and {Loveday}, Jon and {Lucinio}, Rich and {Lupton}, Robert H. and {MacKinnon}, Bryan and {Mannery}, Edward J. and {Mantsch}, P.~M. and {Margon}, Bruce and {McGehee}, Peregrine and {McKay}, Timothy A. and {Meiksin}, Avery and {Merelli}, Aronne and {Monet}, David G. and {Munn}, Jeffrey A. and {Narayanan}, Vijay K. and {Nash}, Thomas and {Neilsen}, Eric and {Neswold}, Rich and {Newberg}, Heidi Jo and {Nichol}, R.~C. and {Nicinski}, Tom and {Nonino}, Mario and {Okada}, Norio and {Okamura}, Sadanori and {Ostriker}, Jeremiah P. and {Owen}, Russell and {Pauls}, A. George and {Peoples}, John and {Peterson}, R.~L. and {Petravick}, Donald and {Pier}, Jeffrey R. and {Pope}, Adrian and {Pordes}, Ruth and {Prosapio}, Angela and {Rechenmacher}, Ron and {Quinn}, Thomas R. and {Richards}, Gordon T. and {Richmond}, Michael W. and {Rivetta}, Claudio H. and {Rockosi}, Constance M. and {Ruthmansdorfer}, Kurt and {Sandford}, Dale and {Schlegel}, David J. and {Schneider}, Donald P. and {Sekiguchi}, Maki and {Sergey}, Gary and {Shimasaku}, Kazuhiro and {Siegmund}, Walter A. and {Smee}, Stephen and {Smith}, J. Allyn and {Snedden}, S. and {Stone}, R. and {Stoughton}, Chris and {Strauss}, Michael A. and {Stubbs}, Christopher and {SubbaRao}, Mark and {Szalay}, Alexander S. and {Szapudi}, Istvan and {Szokoly}, Gyula P. and {Thakar}, Anirudda R. and {Tremonti}, Christy and {Tucker}, Douglas L. and {Uomoto}, Alan and {Vanden Berk}, Dan and {Vogeley}, Michael S. and {Waddell}, Patrick and {Wang}, Shu-i. and {Watanabe}, Masaru and {Weinberg}, David H. and {Yanny}, Brian and {Yasuda}, Naoki and {SDSS Collaboration}},
        title = "{The Sloan Digital Sky Survey: Technical Summary}",
      journal = {\aj},
         year = 2000,
        month = sep,
       volume = {120},
       number = {3},
        pages = {1579-1587},
          doi = {10.1086/301513},
archivePrefix = {arXiv},
       eprint = {astro-ph/0006396},
 primaryClass = {astro-ph},
       adsurl = {https://ui.adsabs.harvard.edu/abs/2000AJ....120.1579Y}
}

@ARTICLE{lsst,
       author = {{Ivezi{\'c}}, {\v{Z}}eljko and {Kahn}, Steven M. and {Tyson}, J. Anthony and {Abel}, Bob and {Acosta}, Emily and {Allsman}, Robyn and {Alonso}, David and {AlSayyad}, Yusra and {Anderson}, Scott F. and {Andrew}, John and {Angel}, James Roger P. and {Angeli}, George Z. and {Ansari}, Reza and {Antilogus}, Pierre and {Araujo}, Constanza and {Armstrong}, Robert and {Arndt}, Kirk T. and {Astier}, Pierre and {Aubourg}, {\'E}ric and {Auza}, Nicole and {Axelrod}, Tim S. and {Bard}, Deborah J. and {Barr}, Jeff D. and {Barrau}, Aurelian and {Bartlett}, James G. and {Bauer}, Amanda E. and {Bauman}, Brian J. and {Baumont}, Sylvain and {Bechtol}, Ellen and {Bechtol}, Keith and {Becker}, Andrew C. and {Becla}, Jacek and {Beldica}, Cristina and {Bellavia}, Steve and {Bianco}, Federica B. and {Biswas}, Rahul and {Blanc}, Guillaume and {Blazek}, Jonathan and {Blandford}, Roger D. and {Bloom}, Josh S. and {Bogart}, Joanne and {Bond}, Tim W. and {Booth}, Michael T. and {Borgland}, Anders W. and {Borne}, Kirk and {Bosch}, James F. and {Boutigny}, Dominique and {Brackett}, Craig A. and {Bradshaw}, Andrew and {Brandt}, William Nielsen and {Brown}, Michael E. and {Bullock}, James S. and {Burchat}, Patricia and {Burke}, David L. and {Cagnoli}, Gianpietro and {Calabrese}, Daniel and {Callahan}, Shawn and {Callen}, Alice L. and {Carlin}, Jeffrey L. and {Carlson}, Erin L. and {Chandrasekharan}, Srinivasan and {Charles-Emerson}, Glenaver and {Chesley}, Steve and {Cheu}, Elliott C. and {Chiang}, Hsin-Fang and {Chiang}, James and {Chirino}, Carol and {Chow}, Derek and {Ciardi}, David R. and {Claver}, Charles F. and {Cohen-Tanugi}, Johann and {Cockrum}, Joseph J. and {Coles}, Rebecca and {Connolly}, Andrew J. and {Cook}, Kem H. and {Cooray}, Asantha and {Covey}, Kevin R. and {Cribbs}, Chris and {Cui}, Wei and {Cutri}, Roc and {Daly}, Philip N. and {Daniel}, Scott F. and {Daruich}, Felipe and {Daubard}, Guillaume and {Daues}, Greg and {Dawson}, William and {Delgado}, Francisco and {Dellapenna}, Alfred and {de Peyster}, Robert and {de Val-Borro}, Miguel and {Digel}, Seth W. and {Doherty}, Peter and {Dubois}, Richard and {Dubois-Felsmann}, Gregory P. and {Durech}, Josef and {Economou}, Frossie and {Eifler}, Tim and {Eracleous}, Michael and {Emmons}, Benjamin L. and {Fausti Neto}, Angelo and {Ferguson}, Henry and {Figueroa}, Enrique and {Fisher-Levine}, Merlin and {Focke}, Warren and {Foss}, Michael D. and {Frank}, James and {Freemon}, Michael D. and {Gangler}, Emmanuel and {Gawiser}, Eric and {Geary}, John C. and {Gee}, Perry and {Geha}, Marla and {Gessner}, Charles J.~B. and {Gibson}, Robert R. and {Gilmore}, D. Kirk and {Glanzman}, Thomas and {Glick}, William and {Goldina}, Tatiana and {Goldstein}, Daniel A. and {Goodenow}, Iain and {Graham}, Melissa L. and {Gressler}, William J. and {Gris}, Philippe and {Guy}, Leanne P. and {Guyonnet}, Augustin and {Haller}, Gunther and {Harris}, Ron and {Hascall}, Patrick A. and {Haupt}, Justine and {Hernandez}, Fabio and {Herrmann}, Sven and {Hileman}, Edward and {Hoblitt}, Joshua and {Hodgson}, John A. and {Hogan}, Craig and {Howard}, James D. and {Huang}, Dajun and {Huffer}, Michael E. and {Ingraham}, Patrick and {Innes}, Walter R. and {Jacoby}, Suzanne H. and {Jain}, Bhuvnesh and {Jammes}, Fabrice and {Jee}, M. James and {Jenness}, Tim and {Jernigan}, Garrett and {Jevremovi{\'c}}, Darko and {Johns}, Kenneth and {Johnson}, Anthony S. and {Johnson}, Margaret W.~G. and {Jones}, R. Lynne and {Juramy-Gilles}, Claire and {Juri{\'c}}, Mario and {Kalirai}, Jason S. and {Kallivayalil}, Nitya J. and {Kalmbach}, Bryce and {Kantor}, Jeffrey P. and {Karst}, Pierre and {Kasliwal}, Mansi M. and {Kelly}, Heather and {Kessler}, Richard and {Kinnison}, Veronica and {Kirkby}, David and {Knox}, Lloyd and {Kotov}, Ivan V. and {Krabbendam}, Victor L. and {Krughoff}, K. Simon and {Kub{\'a}nek}, Petr and {Kuczewski}, John and {Kulkarni}, Shri and {Ku}, John and {Kurita}, Nadine R. and {Lage}, Craig S. and {Lambert}, Ron and {Lange}, Travis and {Langton}, J. Brian and {Le Guillou}, Laurent and {Levine}, Deborah and {Liang}, Ming and {Lim}, Kian-Tat and {Lintott}, Chris J. and {Long}, Kevin E. and {Lopez}, Margaux and {Lotz}, Paul J. and {Lupton}, Robert H. and {Lust}, Nate B. and {MacArthur}, Lauren A. and {Mahabal}, Ashish and {Mandelbaum}, Rachel and {Markiewicz}, Thomas W. and {Marsh}, Darren S. and {Marshall}, Philip J. and {Marshall}, Stuart and {May}, Morgan and {McKercher}, Robert and {McQueen}, Michelle and {Meyers}, Joshua and {Migliore}, Myriam and {Miller}, Michelle and {Mills}, David J.},
        title = "{LSST: From Science Drivers to Reference Design and Anticipated Data Products}",
      journal = {\apj},
         year = 2019,
        month = mar,
       volume = {873},
       number = {2},
          eid = {111},
        pages = {111},
          doi = {10.3847/1538-4357/ab042c},
archivePrefix = {arXiv},
       eprint = {0805.2366},
 primaryClass = {astro-ph},
       adsurl = {https://ui.adsabs.harvard.edu/abs/2019ApJ...873..111I}
}

@PROCEEDINGS{hipparcos,
        title = "{The HIPPARCOS and TYCHO catalogues. Astrometric and photometric star catalogues derived from the ESA HIPPARCOS Space Astrometry Mission}",
    booktitle = {ESA Special Publication},
         year = 1997,
       editor = {{ESA}},
       series = {ESA Special Publication},
       volume = {1200},
        month = jan,
       adsurl = {https://ui.adsabs.harvard.edu/abs/1997ESASP1200.....E}
}

@article{astropy:2013,
Adsurl = {http://adsabs.harvard.edu/abs/2013A%26A...558A..33A},
Archiveprefix = {arXiv},
Author = {{Astropy Collaboration} and {Robitaille}, T.~P. and {Tollerud}, E.~J. and {Greenfield}, P. and {Droettboom}, M. and {Bray}, E. and {Aldcroft}, T. and {Davis}, M. and {Ginsburg}, A. and {Price-Whelan}, A.~M. and {Kerzendorf}, W.~E. and {Conley}, A. and {Crighton}, N. and {Barbary}, K. and {Muna}, D. and {Ferguson}, H. and {Grollier}, F. and {Parikh}, M.~M. and {Nair}, P.~H. and {Unther}, H.~M. and {Deil}, C. and {Woillez}, J. and {Conseil}, S. and {Kramer}, R. and {Turner}, J.~E.~H. and {Singer}, L. and {Fox}, R. and {Weaver}, B.~A. and {Zabalza}, V. and {Edwards}, Z.~I. and {Azalee Bostroem}, K. and {Burke}, D.~J. and {Casey}, A.~R. and {Crawford}, S.~M. and {Dencheva}, N. and {Ely}, J. and {Jenness}, T. and {Labrie}, K. and {Lim}, P.~L. and {Pierfederici}, F. and {Pontzen}, A. and {Ptak}, A. and {Refsdal}, B. and {Servillat}, M. and {Streicher}, O.},
Doi = {10.1051/0004-6361/201322068},
Eid = {A33},
Eprint = {1307.6212},
Journal = {\aap},
Month = oct,
Pages = {A33},
Primaryclass = {astro-ph.IM},
Title = {{Astropy: A community Python package for astronomy}},
Volume = 558,
Year = 2013}

@ARTICLE{astropy:2018,
       author = {{Astropy Collaboration} and {Price-Whelan}, A.~M. and
         {Sip{\H{o}}cz}, B.~M. and {G{\"u}nther}, H.~M. and {Lim}, P.~L. and
         {Crawford}, S.~M. and {Conseil}, S. and {Shupe}, D.~L. and
         {Craig}, M.~W. and {Dencheva}, N. and {Ginsburg}, A. and {Vand
        erPlas}, J.~T. and {Bradley}, L.~D. and {P{\'e}rez-Su{\'a}rez}, D. and
         {de Val-Borro}, M. and {Aldcroft}, T.~L. and {Cruz}, K.~L. and
         {Robitaille}, T.~P. and {Tollerud}, E.~J. and {Ardelean}, C. and
         {Babej}, T. and {Bach}, Y.~P. and {Bachetti}, M. and {Bakanov}, A.~V. and
         {Bamford}, S.~P. and {Barentsen}, G. and {Barmby}, P. and
         {Baumbach}, A. and {Berry}, K.~L. and {Biscani}, F. and {Boquien}, M. and
         {Bostroem}, K.~A. and {Bouma}, L.~G. and {Brammer}, G.~B. and
         {Bray}, E.~M. and {Breytenbach}, H. and {Buddelmeijer}, H. and
         {Burke}, D.~J. and {Calderone}, G. and {Cano Rodr{\'\i}guez}, J.~L. and
         {Cara}, M. and {Cardoso}, J.~V.~M. and {Cheedella}, S. and {Copin}, Y. and
         {Corrales}, L. and {Crichton}, D. and {D'Avella}, D. and {Deil}, C. and
         {Depagne}, {\'E}. and {Dietrich}, J.~P. and {Donath}, A. and
         {Droettboom}, M. and {Earl}, N. and {Erben}, T. and {Fabbro}, S. and
         {Ferreira}, L.~A. and {Finethy}, T. and {Fox}, R.~T. and
         {Garrison}, L.~H. and {Gibbons}, S.~L.~J. and {Goldstein}, D.~A. and
         {Gommers}, R. and {Greco}, J.~P. and {Greenfield}, P. and
         {Groener}, A.~M. and {Grollier}, F. and {Hagen}, A. and {Hirst}, P. and
         {Homeier}, D. and {Horton}, A.~J. and {Hosseinzadeh}, G. and {Hu}, L. and
         {Hunkeler}, J.~S. and {Ivezi{\'c}}, {\v{Z}}. and {Jain}, A. and
         {Jenness}, T. and {Kanarek}, G. and {Kendrew}, S. and {Kern}, N.~S. and
         {Kerzendorf}, W.~E. and {Khvalko}, A. and {King}, J. and {Kirkby}, D. and
         {Kulkarni}, A.~M. and {Kumar}, A. and {Lee}, A. and {Lenz}, D. and
         {Littlefair}, S.~P. and {Ma}, Z. and {Macleod}, D.~M. and
         {Mastropietro}, M. and {McCully}, C. and {Montagnac}, S. and
         {Morris}, B.~M. and {Mueller}, M. and {Mumford}, S.~J. and {Muna}, D. and
         {Murphy}, N.~A. and {Nelson}, S. and {Nguyen}, G.~H. and
         {Ninan}, J.~P. and {N{\"o}the}, M. and {Ogaz}, S. and {Oh}, S. and
         {Parejko}, J.~K. and {Parley}, N. and {Pascual}, S. and {Patil}, R. and
         {Patil}, A.~A. and {Plunkett}, A.~L. and {Prochaska}, J.~X. and
         {Rastogi}, T. and {Reddy Janga}, V. and {Sabater}, J. and
         {Sakurikar}, P. and {Seifert}, M. and {Sherbert}, L.~E. and
         {Sherwood-Taylor}, H. and {Shih}, A.~Y. and {Sick}, J. and
         {Silbiger}, M.~T. and {Singanamalla}, S. and {Singer}, L.~P. and
         {Sladen}, P.~H. and {Sooley}, K.~A. and {Sornarajah}, S. and
         {Streicher}, O. and {Teuben}, P. and {Thomas}, S.~W. and
         {Tremblay}, G.~R. and {Turner}, J.~E.~H. and {Terr{\'o}n}, V. and
         {van Kerkwijk}, M.~H. and {de la Vega}, A. and {Watkins}, L.~L. and
         {Weaver}, B.~A. and {Whitmore}, J.~B. and {Woillez}, J. and
         {Zabalza}, V. and {Astropy Contributors}},
        title = "{The Astropy Project: Building an Open-science Project and Status of the v2.0 Core Package}",
      journal = {\aj},
         year = 2018,
        month = sep,
       volume = {156},
       number = {3},
          eid = {123},
        pages = {123},
          doi = {10.3847/1538-3881/aabc4f},
archivePrefix = {arXiv},
       eprint = {1801.02634},
 primaryClass = {astro-ph.IM},
       adsurl = {https://ui.adsabs.harvard.edu/abs/2018AJ....156..123A}
}

@ARTICLE{astropy:2022,
       author = {{Astropy Collaboration} and {Price-Whelan}, Adrian M. and {Lim}, Pey Lian and {Earl}, Nicholas and {Starkman}, Nathaniel and {Bradley}, Larry and {Shupe}, David L. and {Patil}, Aarya A. and {Corrales}, Lia and {Brasseur}, C.~E. and {N{"o}the}, Maximilian and {Donath}, Axel and {Tollerud}, Erik and {Morris}, Brett M. and {Ginsburg}, Adam and {Vaher}, Eero and {Weaver}, Benjamin A. and {Tocknell}, James and {Jamieson}, William and {van Kerkwijk}, Marten H. and {Robitaille}, Thomas P. and {Merry}, Bruce and {Bachetti}, Matteo and {G{"u}nther}, H. Moritz and {Aldcroft}, Thomas L. and {Alvarado-Montes}, Jaime A. and {Archibald}, Anne M. and {B{'o}di}, Attila and {Bapat}, Shreyas and {Barentsen}, Geert and {Baz{'a}n}, Juanjo and {Biswas}, Manish and {Boquien}, M{'e}d{'e}ric and {Burke}, D.~J. and {Cara}, Daria and {Cara}, Mihai and {Conroy}, Kyle E. and {Conseil}, Simon and {Craig}, Matthew W. and {Cross}, Robert M. and {Cruz}, Kelle L. and {D'Eugenio}, Francesco and {Dencheva}, Nadia and {Devillepoix}, Hadrien A.~R. and {Dietrich}, J{"o}rg P. and {Eigenbrot}, Arthur Davis and {Erben}, Thomas and {Ferreira}, Leonardo and {Foreman-Mackey}, Daniel and {Fox}, Ryan and {Freij}, Nabil and {Garg}, Suyog and {Geda}, Robel and {Glattly}, Lauren and {Gondhalekar}, Yash and {Gordon}, Karl D. and {Grant}, David and {Greenfield}, Perry and {Groener}, Austen M. and {Guest}, Steve and {Gurovich}, Sebastian and {Handberg}, Rasmus and {Hart}, Akeem and {Hatfield-Dodds}, Zac and {Homeier}, Derek and {Hosseinzadeh}, Griffin and {Jenness}, Tim and {Jones}, Craig K. and {Joseph}, Prajwel and {Kalmbach}, J. Bryce and {Karamehmetoglu}, Emir and {Ka{l}uszy{'n}ski}, Miko{l}aj and {Kelley}, Michael S.~P. and {Kern}, Nicholas and {Kerzendorf}, Wolfgang E. and {Koch}, Eric W. and {Kulumani}, Shankar and {Lee}, Antony and {Ly}, Chun and {Ma}, Zhiyuan and {MacBride}, Conor and {Maljaars}, Jakob M. and {Muna}, Demitri and {Murphy}, N.~A. and {Norman}, Henrik and {O'Steen}, Richard and {Oman}, Kyle A. and {Pacifici}, Camilla and {Pascual}, Sergio and {Pascual-Granado}, J. and {Patil}, Rohit R. and {Perren}, Gabriel I. and {Pickering}, Timothy E. and {Rastogi}, Tanuj and {Roulston}, Benjamin R. and {Ryan}, Daniel F. and {Rykoff}, Eli S. and {Sabater}, Jose and {Sakurikar}, Parikshit and {Salgado}, Jes{'u}s and {Sanghi}, Aniket and {Saunders}, Nicholas and {Savchenko}, Volodymyr and {Schwardt}, Ludwig and {Seifert-Eckert}, Michael and {Shih}, Albert Y. and {Jain}, Anany Shrey and {Shukla}, Gyanendra and {Sick}, Jonathan and {Simpson}, Chris and {Singanamalla}, Sudheesh and {Singer}, Leo P. and {Singhal}, Jaladh and {Sinha}, Manodeep and {Sip{H{o}}cz}, Brigitta M. and {Spitler}, Lee R. and {Stansby}, David and {Streicher}, Ole and {{{S}}umak}, Jani and {Swinbank}, John D. and {Taranu}, Dan S. and {Tewary}, Nikita and {Tremblay}, Grant R. and {Val-Borro}, Miguel de and {Van Kooten}, Samuel J. and {Vasovi{'c}}, Zlatan and {Verma}, Shresth and {de Miranda Cardoso}, Jos{'e} Vin{'i}cius and {Williams}, Peter K.~G. and {Wilson}, Tom J. and {Winkel}, Benjamin and {Wood-Vasey}, W.~M. and {Xue}, Rui and {Yoachim}, Peter and {Zhang}, Chen and {Zonca}, Andrea and {Astropy Project Contributors}},
        title = "{The Astropy Project: Sustaining and Growing a Community-oriented Open-source Project and the Latest Major Release (v5.0) of the Core Package}",
      journal = {\apj},
         year = 2022,
        month = aug,
       volume = {935},
       number = {2},
          eid = {167},
        pages = {167},
          doi = {10.3847/1538-4357/ac7c74},
archivePrefix = {arXiv},
       eprint = {2206.14220},
 primaryClass = {astro-ph.IM},
       adsurl = {https://ui.adsabs.harvard.edu/abs/2022ApJ...935..167A}
}

@article{astrometry,
  author = {Lang, D. and Hogg, D. W. and Mierle, K. and Blanton, M. and Roweis, S.},
  title = {Astrometry.net: Blind astrometric calibration of arbitrary astronomical images},
  journal = {The Astronomical Journal},
  volume = {139},
  pages = {1782–1800},
  year = {2010},
  doi = {10.1088/0004-6256/139/5/1782}
}

@article{photutils,
  author       = {Larry Bradley and
                  Brigitta Sip{\H o}cz and
                  Thomas Robitaille and
                  Erik Tollerud and
                  Z\`e Vin{\'{\i}}cius and
                  Christoph Deil and
                  Kyle Barbary and
                  Tom J Wilson and
                  Ivo Busko and
                  Axel Donath and
                  Hans Moritz G{\"u}nther and
                  Mihai Cara and
                  P. L. Lim and
                  Sebastian Me{\ss}linger and
                  Zach Burnett and
                  Simon Conseil and
                  Michael Droettboom and
                  Azalee Bostroem and
                  E. M. Bray and
                  Lars Andersen Bratholm and
                  William Jamieson and
                  Adam Ginsburg and
                  Geert Barentsen and
                  Matt Craig and
                  Sergio Pascual and
                  Shivangee Rathi and
                  Marshall Perrin and
                  Brett M. Morris},
  title        = {astropy/photutils: 2.2.0},
  month        = feb,
  year         = 2025,
  journal      = {Zenodo},
  publisher    = {Zenodo},
  version      = {2.2.0},
  doi          = {10.5281/zenodo.14889440},
  url          = {https://doi.org/10.5281/zenodo.14889440},
  swhid        = {swh:1:dir:11159107f27a28985192ed1118b1f2055709d093
                   ;origin=https://doi.org/10.5281/zenodo.596036;visi
                   t=swh:1:snp:ae8c4a55d349d43e53cfe9ce92e678fcfe840f
                   3b;anchor=swh:1:rel:0117f67e8888adcdfc85308287dd9c
                   854b466389;path=astropy-photutils-ffb96c5
                  },
}

@ARTICLE{astroquery,
   author = {{Ginsburg}, A. and {Sip{\H o}cz}, B.~M. and {Brasseur}, C.~E. and
	{Cowperthwaite}, P.~S. and {Craig}, M.~W. and {Deil}, C. and
	{Guillochon}, J. and {Guzman}, G. and {Liedtke}, S. and {Lian Lim}, P. and
	{Lockhart}, K.~E. and {Mommert}, M. and {Morris}, B.~M. and
	{Norman}, H. and {Parikh}, M. and {Persson}, M.~V. and {Robitaille}, T.~P. and
	{Segovia}, J.-C. and {Singer}, L.~P. and {Tollerud}, E.~J. and
	{de Val-Borro}, M. and {Valtchanov}, I. and {Woillez}, J. and
	{The Astroquery collaboration} and {a subset of the astropy collaboration}
	},
    title = "{astroquery: An Astronomical Web-querying Package in Python}",
  journal = {\aj},
archivePrefix = "arXiv",
   eprint = {1901.04520},
 primaryClass = "astro-ph.IM",
     year = 2019,
    month = mar,
   volume = 157,
      eid = {98},
    pages = {98},
      doi = {10.3847/1538-3881/aafc33},
   adsurl = {https://adsabs.harvard.edu/abs/2019AJ....157...98G}
}

@misc{gaia_mag_transform2022,
	author = {{Carrasco}, J. M. and {Bellazzini}, M.},
	title = {Relationships with other photometric systems},
	year = {2022},
	howpublished = {\url{https://gea.esac.esa.int/archive/documentation/GDR3/Data_processing/chap_cu5pho/cu5pho_sec_photSystem/cu5pho_ssec_photRelations.html\#Ch5.T8}},
	note = "[Online; accessed 29-July-2026]"
}

@article{ransac,
  title={Random sample consensus: a paradigm for model fitting with applications to image analysis and automated cartography},
  author={Fischler, Martin A and Bolles, Robert C},
  journal={Communications of the ACM},
  volume={24},
  number={6},
  pages={381--395},
  year={1981},
  publisher={ACM New York, NY, USA}
}

@ARTICLE{Daofind,
       author = {{Stetson}, Peter B.},
        title = "{DAOPHOT: A Computer Program for Crowded-Field Stellar Photometry}",
      journal = {\pasp},
         year = 1987,
        month = mar,
       volume = {99},
        pages = {191},
          doi = {10.1086/131977},
       adsurl = {https://ui.adsabs.harvard.edu/abs/1987PASP...99..191S}
}

@ARTICLE{penelope,
       author = {{Shevchenko}, V.~G. and {Belskaya}, I.~N. and {Krugly}, Yu. N. and {Chiomy}, V.~G. and {Gaftonyuk}, N.~M.},
        title = "{Asteroid Observations at Low Phase Angles. II. 5 Astraea, 75 Eurydike, 77 Frigga, 105 Artemis, 119 Althaea, 124 Alkeste, and 201 Penelope}",
      journal = {\icarus},
         year = 2002,
        month = feb,
       volume = {155},
       number = {2},
        pages = {365-374},
          doi = {10.1006/icar.2001.6651},
       adsurl = {https://ui.adsabs.harvard.edu/abs/2002Icar..155..365S}
}

@ARTICLE{maria,
       author = {{Kim}, M.-J. and {Choi}, Y.-J. and {Moon}, H.-K. and {Byun}, Y.-I. and {Brosch}, N. and {Kaplan}, M. and {Kaynar}, S. and {Uysal}, {\"O}. and {G{\"u}zel}, E. and {Behrend}, R. and {Yoon}, J.-N. and {Mottola}, S. and {Hellmich}, S. and {Hinse}, T.~C. and {Eker}, Z. and {Park}, J.-H.},
        title = "{Rotational Properties of the Maria Asteroid Family}",
      journal = {\aj},
         year = 2014,
        month = mar,
       volume = {147},
       number = {3},
          eid = {56},
        pages = {56},
          doi = {10.1088/0004-6256/147/3/56},
archivePrefix = {arXiv},
       eprint = {1311.5318},
 primaryClass = {astro-ph.EP},
       adsurl = {https://ui.adsabs.harvard.edu/abs/2014AJ....147...56K}
}

@ARTICLE{Gaia2021-astrometry1,
       author = {{Gaia Collaboration} and {Brown}, A.~G.~A. and {Vallenari}, A. and {Prusti}, T. and {de Bruijne}, J.~H.~J. and {Babusiaux}, C. and {Biermann}, M. and {Creevey}, O.~L. and {Evans}, D.~W. and {Eyer}, L. and {Hutton}, A. and {Jansen}, F. and {Jordi}, C. and {Klioner}, S.~A. and {Lammers}, U. and {Lindegren}, L. and {Luri}, X. and {Mignard}, F. and {Panem}, C. and {Pourbaix}, D. and {Randich}, S. and {Sartoretti}, P. and {Soubiran}, C. and {Walton}, N.~A. and {Arenou}, F. and {Bailer-Jones}, C.~A.~L. and {Bastian}, U. and {Cropper}, M. and {Drimmel}, R. and {Katz}, D. and {Lattanzi}, M.~G. and {van Leeuwen}, F. and {Bakker}, J. and {Cacciari}, C. and {Casta{\~n}eda}, J. and {De Angeli}, F. and {Ducourant}, C. and {Fabricius}, C. and {Fouesneau}, M. and {Fr{\'e}mat}, Y. and {Guerra}, R. and {Guerrier}, A. and {Guiraud}, J. and {Jean-Antoine Piccolo}, A. and {Masana}, E. and {Messineo}, R. and {Mowlavi}, N. and {Nicolas}, C. and {Nienartowicz}, K. and {Pailler}, F. and {Panuzzo}, P. and {Riclet}, F. and {Roux}, W. and {Seabroke}, G.~M. and {Sordo}, R. and {Tanga}, P. and {Th{\'e}venin}, F. and {Gracia-Abril}, G. and {Portell}, J. and {Teyssier}, D. and {Altmann}, M. and {Andrae}, R. and {Bellas-Velidis}, I. and {Benson}, K. and {Berthier}, J. and {Blomme}, R. and {Brugaletta}, E. and {Burgess}, P.~W. and {Busso}, G. and {Carry}, B. and {Cellino}, A. and {Cheek}, N. and {Clementini}, G. and {Damerdji}, Y. and {Davidson}, M. and {Delchambre}, L. and {Dell'Oro}, A. and {Fern{\'a}ndez-Hern{\'a}ndez}, J. and {Galluccio}, L. and {Garc{\'\i}a-Lario}, P. and {Garcia-Reinaldos}, M. and {Gonz{\'a}lez-N{\'u}{\~n}ez}, J. and {Gosset}, E. and {Haigron}, R. and {Halbwachs}, J.-L. and {Hambly}, N.~C. and {Harrison}, D.~L. and {Hatzidimitriou}, D. and {Heiter}, U. and {Hern{\'a}ndez}, J. and {Hestroffer}, D. and {Hodgkin}, S.~T. and {Holl}, B. and {Jan{\ss}en}, K. and {Jevardat de Fombelle}, G. and {Jordan}, S. and {Krone-Martins}, A. and {Lanzafame}, A.~C. and {L{\"o}ffler}, W. and {Lorca}, A. and {Manteiga}, M. and {Marchal}, O. and {Marrese}, P.~M. and {Moitinho}, A. and {Mora}, A. and {Muinonen}, K. and {Osborne}, P. and {Pancino}, E. and {Pauwels}, T. and {Petit}, J.-M. and {Recio-Blanco}, A. and {Richards}, P.~J. and {Riello}, M. and {Rimoldini}, L. and {Robin}, A.~C. and {Roegiers}, T. and {Rybizki}, J. and {Sarro}, L.~M. and {Siopis}, C. and {Smith}, M. and {Sozzetti}, A. and {Ulla}, A. and {Utrilla}, E. and {van Leeuwen}, M. and {van Reeven}, W. and {Abbas}, U. and {Abreu Aramburu}, A. and {Accart}, S. and {Aerts}, C. and {Aguado}, J.~J. and {Ajaj}, M. and {Altavilla}, G. and {{\'A}lvarez}, M.~A. and {{\'A}lvarez Cid-Fuentes}, J. and {Alves}, J. and {Anderson}, R.~I. and {Anglada Varela}, E. and {Antoja}, T. and {Audard}, M. and {Baines}, D. and {Baker}, S.~G. and {Balaguer-N{\'u}{\~n}ez}, L. and {Balbinot}, E. and {Balog}, Z. and {Barache}, C. and {Barbato}, D. and {Barros}, M. and {Barstow}, M.~A. and {Bartolom{\'e}}, S. and {Bassilana}, J.-L. and {Bauchet}, N. and {Baudesson-Stella}, A. and {Becciani}, U. and {Bellazzini}, M. and {Bernet}, M. and {Bertone}, S. and {Bianchi}, L. and {Blanco-Cuaresma}, S. and {Boch}, T. and {Bombrun}, A. and {Bossini}, D. and {Bouquillon}, S. and {Bragaglia}, A. and {Bramante}, L. and {Breedt}, E. and {Bressan}, A. and {Brouillet}, N. and {Bucciarelli}, B. and {Burlacu}, A. and {Busonero}, D. and {Butkevich}, A.~G. and {Buzzi}, R. and {Caffau}, E. and {Cancelliere}, R. and {C{\'a}novas}, H. and {Cantat-Gaudin}, T. and {Carballo}, R. and {Carlucci}, T. and {Carnerero}, M.~I. and {Carrasco}, J.~M. and {Casamiquela}, L. and {Castellani}, M. and {Castro-Ginard}, A. and {Castro Sampol}, P. and {Chaoul}, L. and {Charlot}, P. and {Chemin}, L. and {Chiavassa}, A. and {Cioni}, M.-R.~L. and {Comoretto}, G. and {Cooper}, W.~J. and {Cornez}, T. and {Cowell}, S. and {Crifo}, F. and {Crosta}, M. and {Crowley}, C. and {Dafonte}, C. and {Dapergolas}, A. and {David}, M. and {David}, P.},
        title = "{Gaia Early Data Release 3. Summary of the contents and survey properties}",
      journal = {\aap},
         year = 2021,
        month = may,
       volume = {649},
          eid = {A1},
        pages = {A1},
          doi = {10.1051/0004-6361/202039657},
archivePrefix = {arXiv},
       eprint = {2012.01533},
 primaryClass = {astro-ph.GA},
       adsurl = {https://ui.adsabs.harvard.edu/abs/2021A&A...649A...1G}
}

@ARTICLE{Gaia-astrometry2,
       author = {{Lindegren}, L. and {Klioner}, S.~A. and {Hern{\'a}ndez}, J. and {Bombrun}, A. and {Ramos-Lerate}, M. and {Steidelm{\"u}ller}, H. and {Bastian}, U. and {Biermann}, M. and {de Torres}, A. and {Gerlach}, E. and {Geyer}, R. and {Hilger}, T. and {Hobbs}, D. and {Lammers}, U. and {McMillan}, P.~J. and {Stephenson}, C.~A. and {Casta{\~n}eda}, J. and {Davidson}, M. and {Fabricius}, C. and {Gracia-Abril}, G. and {Portell}, J. and {Rowell}, N. and {Teyssier}, D. and {Torra}, F. and {Bartolom{\'e}}, S. and {Clotet}, M. and {Garralda}, N. and {Gonz{\'a}lez-Vidal}, J.~J. and {Torra}, J. and {Abbas}, U. and {Altmann}, M. and {Anglada Varela}, E. and {Balaguer-N{\'u}{\~n}ez}, L. and {Balog}, Z. and {Barache}, C. and {Becciani}, U. and {Bernet}, M. and {Bertone}, S. and {Bianchi}, L. and {Bouquillon}, S. and {Brown}, A.~G.~A. and {Bucciarelli}, B. and {Busonero}, D. and {Butkevich}, A.~G. and {Buzzi}, R. and {Cancelliere}, R. and {Carlucci}, T. and {Charlot}, P. and {Cioni}, M.-R.~L. and {Crosta}, M. and {Crowley}, C. and {del Peloso}, E.~F. and {del Pozo}, E. and {Drimmel}, R. and {Esquej}, P. and {Fienga}, A. and {Fraile}, E. and {Gai}, M. and {Garcia-Reinaldos}, M. and {Guerra}, R. and {Hambly}, N.~C. and {Hauser}, M. and {Jan{\ss}en}, K. and {Jordan}, S. and {Kostrzewa-Rutkowska}, Z. and {Lattanzi}, M.~G. and {Liao}, S. and {Licata}, E. and {Lister}, T.~A. and {L{\"o}ffler}, W. and {Marchant}, J.~M. and {Masip}, A. and {Mignard}, F. and {Mints}, A. and {Molina}, D. and {Mora}, A. and {Morbidelli}, R. and {Murphy}, C.~P. and {Pagani}, C. and {Panuzzo}, P. and {Pe{\~n}alosa Esteller}, X. and {Poggio}, E. and {Re Fiorentin}, P. and {Riva}, A. and {Sagrist{\`a} Sell{\'e}s}, A. and {Sanchez Gimenez}, V. and {Sarasso}, M. and {Sciacca}, E. and {Siddiqui}, H.~I. and {Smart}, R.~L. and {Souami}, D. and {Spagna}, A. and {Steele}, I.~A. and {Taris}, F. and {Utrilla}, E. and {van Reeven}, W. and {Vecchiato}, A.},
        title = "{Gaia Early Data Release 3. The astrometric solution}",
      journal = {\aap},
         year = 2021,
        month = may,
       volume = {649},
          eid = {A2},
        pages = {A2},
          doi = {10.1051/0004-6361/202039709},
archivePrefix = {arXiv},
       eprint = {2012.03380},
 primaryClass = {astro-ph.IM},
       adsurl = {https://ui.adsabs.harvard.edu/abs/2021A&A...649A...2L}
}

@ARTICLE{v523cas1,
       author = {{Samec}, Ronald G. and {Faulkner}, Danny R. and {Williams}, David B.},
        title = "{The Physical Nature and Orbital Behavior of V523 Cassiopeiae}",
      journal = {\aj},
         year = 2004,
        month = dec,
       volume = {128},
       number = {6},
        pages = {2997-3004},
          doi = {10.1086/426357},
       adsurl = {https://ui.adsabs.harvard.edu/abs/2004AJ....128.2997S}
}

@ARTICLE{v523cas2,
       author = {{Mohammadi}, Mahya and {Abedi}, Abbas and {Riazi}, Nematollah},
        title = "{Period and light-curve study of the contact eclipsing binary V523 Cas}",
      journal = {\na},
         year = 2016,
        month = apr,
       volume = {44},
        pages = {78-85},
          doi = {10.1016/j.newast.2015.10.001},
       adsurl = {https://ui.adsabs.harvard.edu/abs/2016NewA...44...78M}
}

@ARTICLE{v523cas3,
       author = {{K{\"o}se}, O. and {Keskin}, V. and {Yakut}, K.},
        title = "{Absolute dimensions of the close binary systems V453 Monocerotis and V523 Cassiopeiae}",
      journal = {\apss},
         year = 2009,
        month = sep,
       volume = {323},
       number = {1},
        pages = {75-81},
          doi = {10.1007/s10509-009-0048-0},
       adsurl = {https://ui.adsabs.harvard.edu/abs/2009Ap&SS.323...75K}
}

@ARTICLE{edps,
       author = {{Freudling}, W. and {Zampieri}, S. and {Coccato}, L. and {Podgorski}, S. and {Romaniello}, M. and {Modigliani}, A. and {Pritchard}, J.},
        title = "{Adaptive data reduction workflows for astronomy: The ESO Data Processing System (EDPS)}",
      journal = {\aap},
         year = 2024,
        month = jan,
       volume = {681},
          eid = {A93},
        pages = {A93},
          doi = {10.1051/0004-6361/202347651},
archivePrefix = {arXiv},
       eprint = {2311.03822},
 primaryClass = {astro-ph.IM},
       adsurl = {https://ui.adsabs.harvard.edu/abs/2024A&A...681A..93F}
}

@ARTICLE{esoreflex,
       author = {{Freudling}, W. and {Romaniello}, M. and {Bramich}, D.~M. and {Ballester}, P. and {Forchi}, V. and {Garc{\'\i}a-Dabl{\'o}}, C.~E. and {Moehler}, S. and {Neeser}, M.~J.},
        title = "{Automated data reduction workflows for astronomy. The ESO Reflex environment}",
      journal = {\aap},
         year = 2013,
        month = nov,
       volume = {559},
          eid = {A96},
        pages = {A96},
          doi = {10.1051/0004-6361/201322494},
archivePrefix = {arXiv},
       eprint = {1311.5411},
 primaryClass = {astro-ph.IM},
       adsurl = {https://ui.adsabs.harvard.edu/abs/2013A&A...559A..96F}
}

@ARTICLE{dragons,
       author = {{Labrie}, K. and {Simpson}, C. and {Cardenes}, R. and {Turner}, J. and {Soraisam}, M. and {Quint}, B. and {Oberdorf}, O. and {Placco}, V.~M. and {Berke}, D. and {Smirnova}, O. and {Conseil}, S. and {Vacca}, W.~D. and {Thomas-Osip}, J.},
        title = "{DRAGONS-A Quick Overview}",
      journal = {Research Notes of the American Astronomical Society},
         year = 2023,
        month = oct,
       volume = {7},
       number = {10},
          eid = {214},
        pages = {214},
          doi = {10.3847/2515-5172/ad0044},
archivePrefix = {arXiv},
       eprint = {2310.03048},
 primaryClass = {astro-ph.IM},
       adsurl = {https://ui.adsabs.harvard.edu/abs/2023RNAAS...7..214L}
}

@ARTICLE{panstarrs,
       author = {{Magnier}, Eugene A. and {Chambers}, K.~C. and {Flewelling}, H.~A. and {Hoblitt}, J.~C. and {Huber}, M.~E. and {Price}, P.~A. and {Sweeney}, W.~E. and {Waters}, C.~Z. and {Denneau}, L. and {Draper}, P.~W. and {Hodapp}, K.~W. and {Jedicke}, R. and {Kaiser}, N. and {Kudritzki}, R.-P. and {Metcalfe}, N. and {Stubbs}, C.~W. and {Wainscoat}, R.~J.},
        title = "{The Pan-STARRS Data-processing System}",
      journal = {\apjs},
         year = 2020,
        month = nov,
       volume = {251},
       number = {1},
          eid = {3},
        pages = {3},
          doi = {10.3847/1538-4365/abb829},
archivePrefix = {arXiv},
       eprint = {1612.05240},
 primaryClass = {astro-ph.IM},
       adsurl = {https://ui.adsabs.harvard.edu/abs/2020ApJS..251....3M}
}

@INPROCEEDINGS{iraf:1986,
       author = {{Tody}, Doug},
        title = "{The IRAF Data Reduction and Analysis System}",
    booktitle = {Instrumentation in astronomy VI},
         year = 1986,
       editor = {{Crawford}, David L.},
       series = {Society of Photo-Optical Instrumentation Engineers (SPIE) Conference Series},
       volume = {627},
        month = jan,
        pages = {733},
          doi = {10.1117/12.968154},
       adsurl = {https://ui.adsabs.harvard.edu/abs/1986SPIE..627..733T}
}

@INPROCEEDINGS{iraf:1993,
       author = {{Tody}, Doug},
        title = "{IRAF in the Nineties}",
    booktitle = {Astronomical Data Analysis Software and Systems II},
         year = 1993,
       editor = {{Hanisch}, R.~J. and {Brissenden}, R.~J.~V. and {Barnes}, J.},
       series = {Astronomical Society of the Pacific Conference Series},
       volume = {52},
        month = jan,
        pages = {173},
       adsurl = {https://ui.adsabs.harvard.edu/abs/1993ASPC...52..173T}
}

@ARTICLE{fits,
       author = {{Wells}, D.~C. and {Greisen}, E.~W. and {Harten}, R.~H.},
        title = "{FITS - a Flexible Image Transport System}",
      journal = {\aaps},
         year = 1981,
        month = jun,
       volume = {44},
        pages = {363},
       adsurl = {https://ui.adsabs.harvard.edu/abs/1981A&AS...44..363W}
}

@INPROCEEDINGS{healpix,
       author = {{G{\'o}rski}, Krzysztof M. and {Banday}, A.~J. and {Hivon}, E. and {Wandelt}, B.~D.},
        title = "{HEALPix --- a Framework for High Resolution, Fast Analysis on the Sphere}",
    booktitle = {Astronomical Data Analysis Software and Systems XI},
         year = 2002,
       editor = {{Bohlender}, David A. and {Durand}, Daniel and {Handley}, Thomas H.},
       series = {Astronomical Society of the Pacific Conference Series},
       volume = {281},
        month = jan,
        pages = {107},
       adsurl = {https://ui.adsabs.harvard.edu/abs/2002ASPC..281..107G}
}

@ARTICLE{erece2023,
       author = {{Erece}, Orhan and {Khamitov}, Irek M. and {Kaplan}, Murat and {Kilic}, Yucel and {Lee}, Hee-Jae and {Kim}, Myung-Jin and {Bikmaev}, Ilfan F. and {Gumerov}, Rustem I. and {Irtuganov}, Eldar N.},
        title = "{Physical properties of the slow-rotating near-Earth asteroid (2059) Baboquivari from one apparition}",
      journal = {\planss},
         year = 2023,
        month = aug,
       volume = {232},
          eid = {105698},
        pages = {105698},
          doi = {10.1016/j.pss.2023.105698},
archivePrefix = {arXiv},
       eprint = {2305.05217},
 primaryClass = {astro-ph.EP},
       adsurl = {https://ui.adsabs.harvard.edu/abs/2023P&SS..23205698E}
}

@ARTICLE{kilic2026,
       author = {{Kilic}, Y. and {Braga-Ribas}, F. and {Pereira}, C.~L. and {Ortiz}, J.~L. and {Sicardy}, B. and {Santos-Sanz}, P. and {Erece}, O. and {Rizos}, J.~L. and {G{\'o}mez-Lim{\'o}n}, J.~M. and {Margoti}, G. and {Souami}, D. and {Morgado}, B. and {Gomes-Junior}, A. and {Catani}, L.~M. and {Desmars}, J. and {Kretlow}, M. and {Rommel}, F. and {Duffard}, R. and {Alvarez-Candal}, A. and {Camargo}, J.~I.~B. and {Kaplan}, M. and {Morales}, N. and {Herald}, D. and {Assafin}, M. and {Benedetti-Rossi}, G. and {Sfair}, R. and {Savalle}, R. and {Arcas-Silva}, J. and {Bernasconi}, L. and {Blank}, T. and {Bonavita}, M. and {Carlson}, N. and {Christophe}, B. and {Colesanti}, C.~A. and {Collins}, M. and {Columba}, G. and {Dunford}, R. and {Dunham}, D.~W. and {Dunham}, J. and {Emilio}, M. and {Ferrante}, W.~G. and {George}, T. and {Hanna}, W. and {Isopi}, G. and {Jones}, R. and {Kenyon}, D.~A. and {Kerr}, S. and {Kouprianov}, V. and {Maley}, P.~D. and {Mallia}, F. and {Mattei}, J. and {Meunier}, M. and {Napoleao}, T. and {Peixoto}, V.~F. and {Pollock}, J. and {Snodgrass}, C. and {Stechina}, A. and {Thomas}, W. and {Venable}, R. and {Viscome}, G.~R. and {Zapparata}, A. and {Bardecker}, J. and {Castro}, N. and {Cebral}, C. and {Chapman}, A. and {Gao}, C. and {Green}, K. and {Guimaraes}, A. and {Jacques}, C. and {Jehin}, E. and {Konishi}, M. and {Leiva}, R. and {Liberato}, L. and {Magliano}, C. and {Mammana}, L.~A. and {Melita}, M. and {Moura}, V. and {Olivera-Cuello}, Y. and {Peiro}, L. and {Spagnotto}, J. and {Stuart}, P.~C. and {Vanzi}, L. and {Wilberger}, A. and {Malacarne}, M.},
        title = "{Constraining the size, shape, and albedo of the large trans-Neptunian object (28978) Ixion with multi-chord stellar occultations}",
      journal = {\aap},
         year = 2026,
        month = mar,
       volume = {707},
        pages = {A70},
          doi = {10.1051/0004-6361/202557970},
       adsurl = {https://ui.adsabs.harvard.edu/abs/2026A&A...707A..70K}
}

@ARTICLE{centroid_quadratic,
       author = {{Vakili}, Mohammadjavad and {Hogg}, David W.},
        title = "{Do fast stellar centroiding methods saturate the Cram{\'e}r-Rao lower bound?}",
      journal = {arXiv e-prints},
         year = 2016,
        month = oct,
          eid = {arXiv:1610.05873},
        pages = {arXiv:1610.05873},
          doi = {10.48550/arXiv.1610.05873},
archivePrefix = {arXiv},
       eprint = {1610.05873},
 primaryClass = {astro-ph.IM},
       adsurl = {https://ui.adsabs.harvard.edu/abs/2016arXiv161005873V}
}

@ARTICLE{ucac4star,
       author = {{Zacharias}, N. and {Finch}, C.~T. and {Girard}, T.~M. and {Henden}, A. and {Bartlett}, J.~L. and {Monet}, D.~G. and {Zacharias}, M.~I.},
        title = "{The Fourth US Naval Observatory CCD Astrograph Catalog (UCAC4)}",
      journal = {\aj},
         year = 2013,
        month = feb,
       volume = {145},
       number = {2},
          eid = {44},
        pages = {44},
          doi = {10.1088/0004-6256/145/2/44},
archivePrefix = {arXiv},
       eprint = {1212.6182},
 primaryClass = {astro-ph.IM},
       adsurl = {https://ui.adsabs.harvard.edu/abs/2013AJ....145...44Z}
}

@ARTICLE{mommert2017,
       author = {{Mommert}, M.},
        title = "{PHOTOMETRYPIPELINE: An automated pipeline for calibrated photometry}",
      journal = {Astronomy and Computing},
         year = 2017,
        month = jan,
       volume = {18},
        pages = {47-53},
          doi = {10.1016/j.ascom.2016.11.002},
archivePrefix = {arXiv},
       eprint = {1702.00834},
 primaryClass = {astro-ph.IM},
       adsurl = {https://ui.adsabs.harvard.edu/abs/2017A&C....18...47M}
}

@misc{karpov2021,
       author = {{Karpov}, Sergey},
        title = "{STDPipe: Simple Transient Detection Pipeline}",
 howpublished = {Astrophysics Source Code Library, record ascl:2112.006},
         year = 2021,
        month = dec,
          eid = {ascl:2112.006},
archivePrefix = {ascl},
       eprint = {2112.006},
       adsurl = {https://ui.adsabs.harvard.edu/abs/2021ascl.soft12006K}
}

\appendix
\nolinenumbers
\section{Additional photometric examples}\label{sec:appendix}

This appendix provides supplementary examples that extend the validation presented in Sect.~\ref{sec:validation}. We first show a frame affected by a transient tracking error during the observations of (201)~Penelope, and then present two additional applications of \texttt{PhoPS}: multi-band stellar photometry of the eclipsing binary V523~Cas and time-series photometry of the asteroid 19184~(1991~TB6). These examples illustrate the behaviour of the pipeline in representative stellar and moving-target datasets, including one non-ideal observing case.

\subsection{Tracking-error case in the (201)~Penelope dataset}

Figure~\ref{fig:tracking_error} shows a frame affected by a short tracking error during the TUG100 observations of (201)~Penelope. The source profiles are visibly distorted, and faint secondary images are present next to sufficiently bright objects. Despite this temporary degradation, the corresponding calibrated photometric measurements remain consistent with the neighbouring points in the final light curves.

 \begin{figure}
\centering
\includegraphics[width=\linewidth]{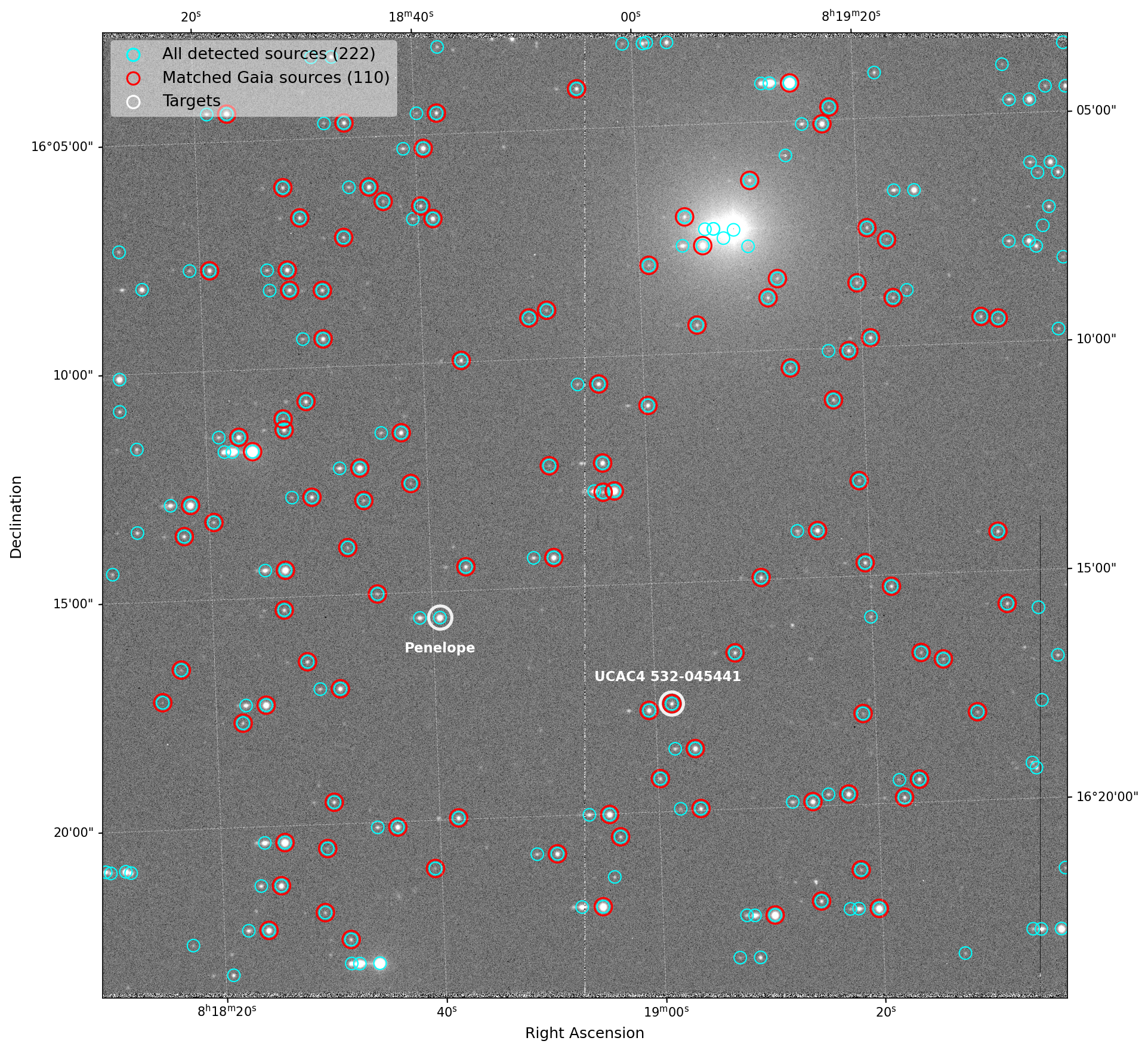}
\caption{Example of a frame affected by a transient tracking error during the observations of (201)~Penelope. The distorted source profiles are visible, and (201)~Penelope and UCAC4 532-045441 are marked.}
\label{fig:tracking_error}
\end{figure}

\subsection{Multi-band photometry of V523~Cas}

To evaluate the performance of \texttt{PhoPS} on stellar time-series photometry, we reduced TUG060 observations of the eclipsing binary V523~Cas, which has a period of about 5.6 h (e.g. \citealt{v523cas1, v523cas3, v523cas2}), obtained in multiple Johnson-Cousins filters. Figure~\ref{fig:V523CAS} shows the resulting light curves. The expected variability pattern is clearly recovered in all bands, indicating that the pipeline can also be successfully applied to multi-band stellar time-series data.

\begin{figure}
\centering
\includegraphics[width=\linewidth]{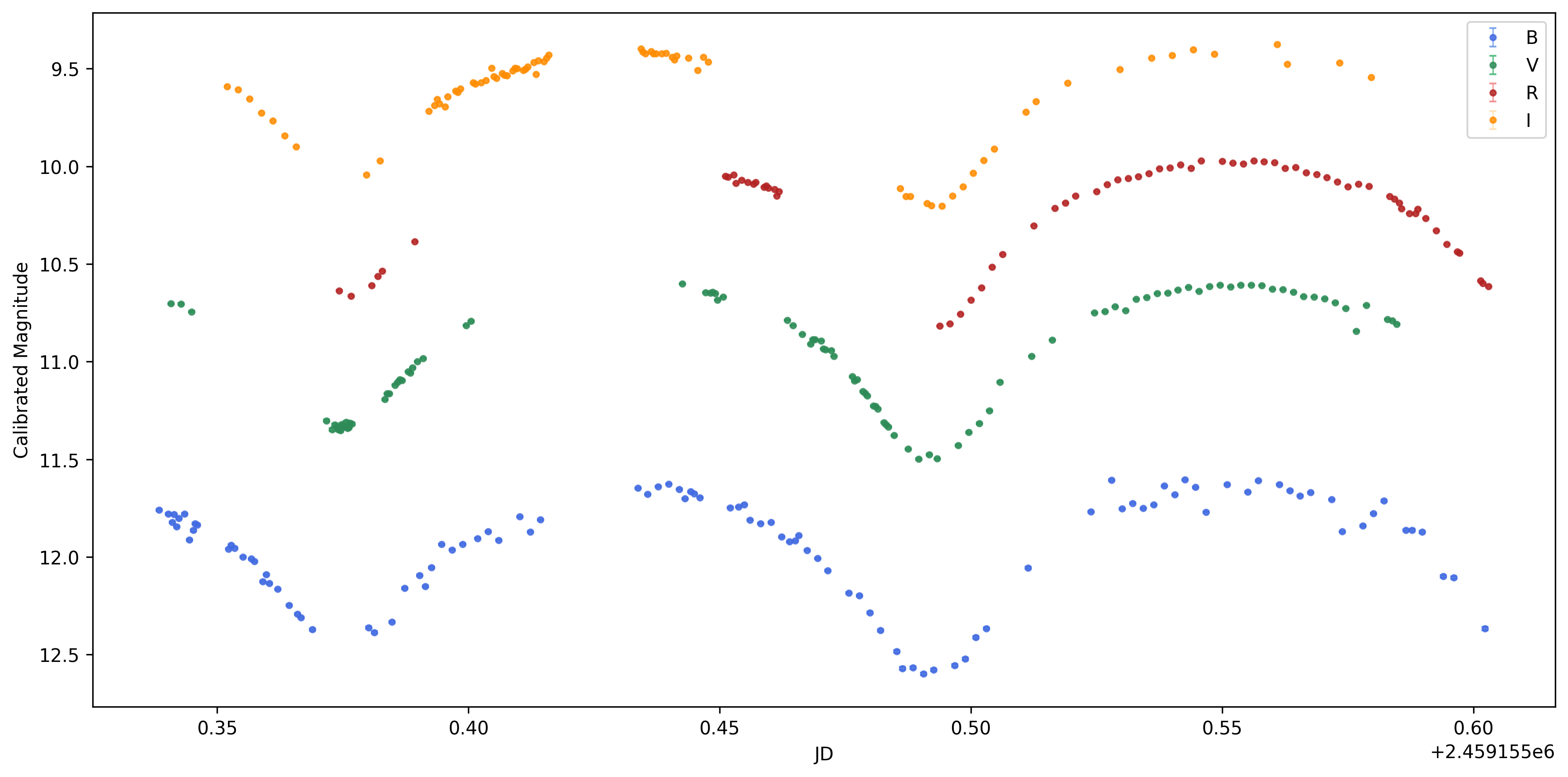}
\caption{Multi-band calibrated light curves of the eclipsing binary V523~Cas derived from TUG060 observations reduced with \texttt{PhoPS}.}
\label{fig:V523CAS}
\end{figure}

\subsection{Photometry of 19184~(1991~TB6)}

As an additional moving target example, we applied \texttt{PhoPS} to TUG100 observations of the asteroid 19184~(1991~TB6). Using the same dataset, \cite{maria} derived a rotational period of 4.99 h for this object. The resulting light curve is shown in Fig.~\ref{fig:19184}. The measured photometric behaviour is consistent with the expected rotational variability of the asteroid and shows that the pipeline can also be applied successfully to independent moving target datasets beyond the main validation case of (201)~Penelope.

\begin{figure}
\centering
\includegraphics[width=\linewidth]{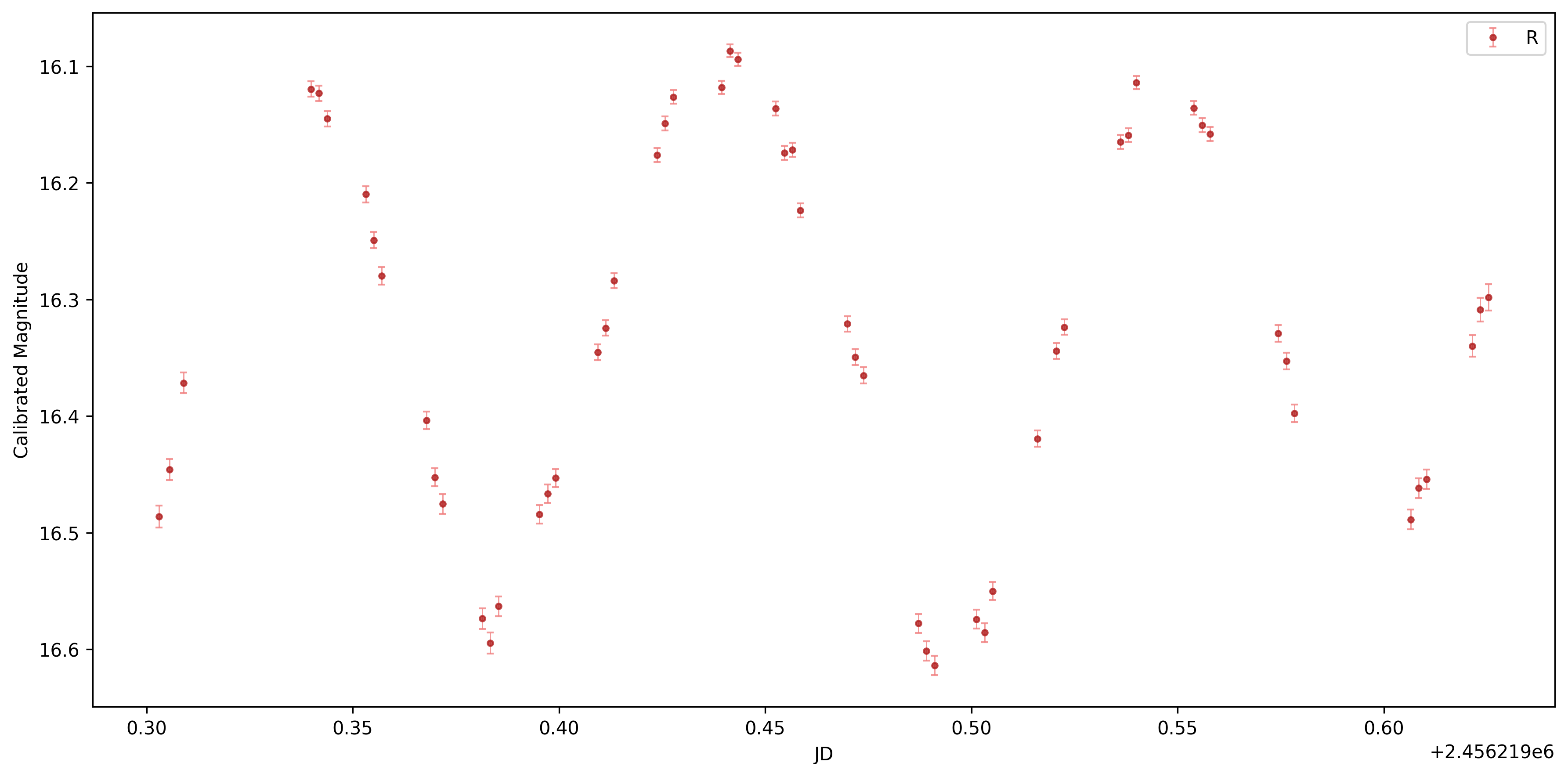}
\caption{Light curve of the asteroid 19184~(1991~TB6) obtained from TUG100 observations and reduced with \texttt{PhoPS}.}
\label{fig:19184}
\end{figure}

\end{document}